\documentclass[aps,prd,twocolumn,showpacs,amsmath,superscriptaddress]{revtex4}
\usepackage{graphicx}  
\usepackage{dcolumn}  
\usepackage{bm}    
\usepackage{amssymb}
\usepackage{epsfig}
\usepackage{slashed}
\usepackage{subfigure}
\usepackage{psfrag}
\usepackage{xcolor}

\def\tev{\ifmmode {\mathrm{~Te\kern -0.1em V}}\else \textrm{Te\kern -0.1em V}\fi}%
\def\gev{\ifmmode {\mathrm{~Ge\kern -0.1em V}}\else \textrm{Ge\kern -0.1em V}\fi}%
\def\pt{\ensuremath{p_{\mathrm{T}}}} % Subscript roman not italic (EE)

\newcommand{\Eslash}{\mbox{$E \kern-0.6em\slash$                }}
\newcommand{\met}{\mbox{\ensuremath{\Eslash_{\kern-0.3emT}\!}                        }}
\newcommand{\mgg}{\mbox{\ensuremath{M_{\gamma\gamma}}}}
\newcommand{\ptgg}{\mbox{\ensuremath{p_{\mathrm{T}}^{\gamma\gamma}}}}
\newcommand{\dphigg}{\mbox{\ensuremath{\Delta\phi_{\gamma\gamma}}}}
\newcommand{\Hgg}{\mbox{\ensuremath{H\to\gamma\gamma}}}
\newcommand{\Hfgg}{\mbox{\ensuremath{H_{\rm f}\to\gamma\gamma}}}
\newcommand{\onn}{\mbox{\ensuremath{O_{\rm NN}}}}
\newcommand{\ptg}{\mbox{\ensuremath{p_{\mathrm{T}}^\gamma}}}
\newcommand{\etag}{\mbox{\ensuremath{\eta^\gamma}}}

\begin{document}

\hspace{5.2in} \mbox{FERMILAB-PUB-13-027-E}

\title{Search for a Higgs boson in diphoton final states with the D0 detector in 9.6 fb$\mathbf{^{-1}}$ of $\mathbf{p\bar{p}}$ collisions at $\mathbf{\sqrt{s} = 1.96~Te\kern -0.1em V}$} 
\affiliation{LAFEX, Centro Brasileiro de Pesquisas F\'{i}sicas, Rio de Janeiro, Brazil}
\affiliation{Universidade do Estado do Rio de Janeiro, Rio de Janeiro, Brazil}
\affiliation{Universidade Federal do ABC, Santo Andr\'e, Brazil}
\affiliation{University of Science and Technology of China, Hefei, People's Republic of China}
\affiliation{Universidad de los Andes, Bogot\'a, Colombia}
\affiliation{Charles University, Faculty of Mathematics and Physics, Center for Particle Physics, Prague, Czech Republic}
\affiliation{Czech Technical University in Prague, Prague, Czech Republic}
\affiliation{Center for Particle Physics, Institute of Physics, Academy of Sciences of the Czech Republic, Prague, Czech Republic}
\affiliation{Universidad San Francisco de Quito, Quito, Ecuador}
\affiliation{LPC, Universit\'e Blaise Pascal, CNRS/IN2P3, Clermont, France}
\affiliation{LPSC, Universit\'e Joseph Fourier Grenoble 1, CNRS/IN2P3, Institut National Polytechnique de Grenoble, Grenoble, France}
\affiliation{CPPM, Aix-Marseille Universit\'e, CNRS/IN2P3, Marseille, France}
\affiliation{LAL, Universit\'e Paris-Sud, CNRS/IN2P3, Orsay, France}
\affiliation{LPNHE, Universit\'es Paris VI and VII, CNRS/IN2P3, Paris, France}
\affiliation{CEA, Irfu, SPP, Saclay, France}
\affiliation{IPHC, Universit\'e de Strasbourg, CNRS/IN2P3, Strasbourg, France}
\affiliation{IPNL, Universit\'e Lyon 1, CNRS/IN2P3, Villeurbanne, France and Universit\'e de Lyon, Lyon, France}
\affiliation{III. Physikalisches Institut A, RWTH Aachen University, Aachen, Germany}
\affiliation{Physikalisches Institut, Universit\"at Freiburg, Freiburg, Germany}
\affiliation{II. Physikalisches Institut, Georg-August-Universit\"at G\"ottingen, G\"ottingen, Germany}
\affiliation{Institut f\"ur Physik, Universit\"at Mainz, Mainz, Germany}
\affiliation{Ludwig-Maximilians-Universit\"at M\"unchen, M\"unchen, Germany}
\affiliation{Fachbereich Physik, Bergische Universit\"at Wuppertal, Wuppertal, Germany}
\affiliation{Panjab University, Chandigarh, India}
\affiliation{Delhi University, Delhi, India}
\affiliation{Tata Institute of Fundamental Research, Mumbai, India}
\affiliation{University College Dublin, Dublin, Ireland}
\affiliation{Korea Detector Laboratory, Korea University, Seoul, Korea}
\affiliation{CINVESTAV, Mexico City, Mexico}
\affiliation{Nikhef, Science Park, Amsterdam, the Netherlands}
\affiliation{Radboud University Nijmegen, Nijmegen, the Netherlands}
\affiliation{Joint Institute for Nuclear Research, Dubna, Russia}
\affiliation{Institute for Theoretical and Experimental Physics, Moscow, Russia}
\affiliation{Moscow State University, Moscow, Russia}
\affiliation{Institute for High Energy Physics, Protvino, Russia}
\affiliation{Petersburg Nuclear Physics Institute, St. Petersburg, Russia}
\affiliation{Instituci\'{o} Catalana de Recerca i Estudis Avan\c{c}ats (ICREA) and Institut de F\'{i}sica d'Altes Energies (IFAE), Barcelona, Spain}
\affiliation{Uppsala University, Uppsala, Sweden}
\affiliation{Lancaster University, Lancaster LA1 4YB, United Kingdom}
\affiliation{Imperial College London, London SW7 2AZ, United Kingdom}
\affiliation{The University of Manchester, Manchester M13 9PL, United Kingdom}
\affiliation{University of Arizona, Tucson, Arizona 85721, USA}
\affiliation{University of California Riverside, Riverside, California 92521, USA}
\affiliation{Florida State University, Tallahassee, Florida 32306, USA}
\affiliation{Fermi National Accelerator Laboratory, Batavia, Illinois 60510, USA}
\affiliation{University of Illinois at Chicago, Chicago, Illinois 60607, USA}
\affiliation{Northern Illinois University, DeKalb, Illinois 60115, USA}
\affiliation{Northwestern University, Evanston, Illinois 60208, USA}
\affiliation{Indiana University, Bloomington, Indiana 47405, USA}
\affiliation{Purdue University Calumet, Hammond, Indiana 46323, USA}
\affiliation{University of Notre Dame, Notre Dame, Indiana 46556, USA}
\affiliation{Iowa State University, Ames, Iowa 50011, USA}
\affiliation{University of Kansas, Lawrence, Kansas 66045, USA}
\affiliation{Louisiana Tech University, Ruston, Louisiana 71272, USA}
\affiliation{Northeastern University, Boston, Massachusetts 02115, USA}
\affiliation{University of Michigan, Ann Arbor, Michigan 48109, USA}
\affiliation{Michigan State University, East Lansing, Michigan 48824, USA}
\affiliation{University of Mississippi, University, Mississippi 38677, USA}
\affiliation{University of Nebraska, Lincoln, Nebraska 68588, USA}
\affiliation{Rutgers University, Piscataway, New Jersey 08855, USA}
\affiliation{Princeton University, Princeton, New Jersey 08544, USA}
\affiliation{State University of New York, Buffalo, New York 14260, USA}
\affiliation{University of Rochester, Rochester, New York 14627, USA}
\affiliation{State University of New York, Stony Brook, New York 11794, USA}
\affiliation{Brookhaven National Laboratory, Upton, New York 11973, USA}
\affiliation{Langston University, Langston, Oklahoma 73050, USA}
\affiliation{University of Oklahoma, Norman, Oklahoma 73019, USA}
\affiliation{Oklahoma State University, Stillwater, Oklahoma 74078, USA}
\affiliation{Brown University, Providence, Rhode Island 02912, USA}
\affiliation{University of Texas, Arlington, Texas 76019, USA}
\affiliation{Southern Methodist University, Dallas, Texas 75275, USA}
\affiliation{Rice University, Houston, Texas 77005, USA}
\affiliation{University of Virginia, Charlottesville, Virginia 22904, USA}
\affiliation{University of Washington, Seattle, Washington 98195, USA}
\author{V.M.~Abazov} \affiliation{Joint Institute for Nuclear Research, Dubna, Russia}
\author{B.~Abbott} \affiliation{University of Oklahoma, Norman, Oklahoma 73019, USA}
\author{B.S.~Acharya} \affiliation{Tata Institute of Fundamental Research, Mumbai, India}
\author{M.~Adams} \affiliation{University of Illinois at Chicago, Chicago, Illinois 60607, USA}
\author{T.~Adams} \affiliation{Florida State University, Tallahassee, Florida 32306, USA}
\author{G.D.~Alexeev} \affiliation{Joint Institute for Nuclear Research, Dubna, Russia}
\author{G.~Alkhazov} \affiliation{Petersburg Nuclear Physics Institute, St. Petersburg, Russia}
\author{A.~Alton$^{a}$} \affiliation{University of Michigan, Ann Arbor, Michigan 48109, USA}
\author{A.~Askew} \affiliation{Florida State University, Tallahassee, Florida 32306, USA}
\author{S.~Atkins} \affiliation{Louisiana Tech University, Ruston, Louisiana 71272, USA}
% ...
%
% visitor_addresses.tex                       11 January 2013 
%  available symbols are:
%  $\ast, \dag, \ddag, \S, \P, $\|$, $\ast\ast$, \dag\dag, \ddag\ddag ,\#
%
\collaboration{The D0 Collaboration\footnote{with visitors from
%{alton}
$^{a}$Augustana College, Sioux Falls, SD, USA,
%{burdin}
$^{b}$The University of Liverpool, Liverpool, UK,
%{garcia-guerra}
$^{c}$UPIITA-IPN, Mexico City, Mexico,
%{grohsjean}
$^{d}$DESY, Hamburg, Germany,
%{partridge}
$^{e}$SLAC, Menlo Park, CA, USA,
%{hesketh}
$^{f}$University College London, London, UK,
%{luna-garcia}
$^{g}$Centro de Investigacion en Computacion - IPN, Mexico City, Mexico,
%{podesta-lerma}
$^{h}$ECFM, Universidad Autonoma de Sinaloa, Culiac\'an, Mexico,
%{santos}
$^{i}$Universidade Estadual Paulista, S\~ao Paulo, Brazil,
%{meyer}
$^{j}$Karlsruher Institut f\"ur Technologie (KIT) - Steinbuch Centre for Computing (SCC)
and
%{patwa}
$^{k}$Office of Science, U.S. Department of Energy, Washington, D.C. 20585, USA.
%{falkowski}
%$^{?}$Laboratoire de Physique Theorique, Orsay, FR,
%{hooper}
%$^{?}$Visitor from Bradley University, Peoria, IL, USA.
%{kozminski}
%$^{?}$}Visitor from Lewis University, Romeoville, IL, USA.
%{weber}
%$^{?}$Universit{\"a}t Bern, Bern, Switzerland.
%{deceased}
%$^{\ddag}$Deceased.
}} \noaffiliation
\vskip 0.25cm
    
\vspace*{0.5cm}   
\date{January 22, 2013}
\begin{abstract}
We present a search for a Higgs boson decaying into a pair of photons based on 9.6 fb$^{-1}$ of $p\bar{p}$ collisions at 
$\sqrt{s} = 1.96\tev$ collected with the D0 detector at the Fermilab Tevatron Collider. The search employs multivariate
techniques to discriminate signal from the non-resonant background and is separately optimized
for a standard model and a fermiophobic Higgs boson. No significant excess of data above the background prediction
is observed and upper limits on the product of the cross section and branching fraction are derived at the 95\% confidence level
as a function of Higgs boson mass. For a standard model Higgs boson with mass of $125\gev$, the  observed (expected) upper limits 
are a factor of 12.8 (8.7) above the standard model prediction. The existence of a fermiophobic Higgs boson with mass in the 100--113$\gev$ 
range is excluded at the 95\% confidence level.
\end{abstract}
\maketitle

\section{Introduction}

Unraveling the mechanism for electroweak symmetry breaking and the generation of mass of
elementary particles has been a priority in experimental particle physics research during the
last decades. In the standard model (SM)~\cite{sm} this is accomplished by introducing a
SU(2) doublet of self-interacting elementary scalars, the ``Higgs field", whose non-zero vacuum expectation
value breaks the electroweak symmetry and generates the mass of the $W$ and $Z$ bosons~\cite{higgs}.
The postulated Yukawa interactions between the fermions and the Higgs field also gives mass to fermions
upon the breaking of the electroweak symmetry. Furthermore, a physical scalar particle appears in the spectrum, the
Higgs boson ($H$), whose mass is not predicted and must be determined experimentally.

Within the SM, indirect constraints from precision electroweak observables~\cite{lepewwg} limit 
the allowed range for the Higgs boson mass ($M_H$) to $M_H<152\gev$ at the 95\% confidence level (CL).
Direct searches at the CERN $e^+e^-$ Collider (LEP)~\cite{lepcombo} set a lower limit of
$M_H>114.4\gev$ at 95\% CL. At hadron colliders the dominant production mechanisms for a SM Higgs 
boson are gluon fusion (GF) ($gg\to H$),  associated production with a $W$ or $Z$ boson ($q\bar{q}^\prime \to VH$, $V=W,Z$), 
and vector boson fusion (VBF) ($VV\to H$). However, the search strategies for a light SM Higgs boson are different at the Fermilab
Tevatron $p\bar{p}$ Collider and at CERN's Large Hadron $pp$ Collider (LHC).

At the Tevatron, the most sensitive SM Higgs boson searches for  $M_H<130\gev$ rely on the $VH$ production mode, with $H\to b\bar{b}$, while 
for $M_H>130\gev$ the main search mode is $gg \to H \to W^+W^-$. The combination of searches at the Tevatron~\cite{tevcombo} 
have resulted in the mass ranges $100<M_H<103\gev$ and $147<M_H<180\gev$ being excluded at the 95\% CL.
In the allowed intermediate mass range an excess is found with a maximum local significance of 3.1 standard deviations (s.d.) at $M_H=125\gev$,
primarily originating from the {\sl VH} ($H\to b\bar{b}$) searches~\cite{tevcombobb}.

At the LHC, the search strategy for $M_H>140\gev$ also capitalizes on the GF production mode, exploiting primarily the 
$H \to W^+W^-$ and $H \to ZZ$ decay modes with leptonic $W$ and $Z$ boson decays. The $\Hgg$ 
decay mode becomes one of the most promising discovery channels at lower $M_H$, despite its small branching fraction of ${\cal B}(\Hgg)\approx 0.2\%$, 
owing to its clean experimental signature of a narrow resonance on top of a smoothly-falling background in the diphoton mass ($\mgg$) spectrum.
Searches for $H \to ZZ^{(*)} \to \ell^+\ell^- \ell^{\prime +} \ell^{\prime -}$ ($\ell, \ell^\prime = e,\mu$) are also
sensitive due to the small background and excellent four-lepton invariant mass resolution.
The most recent searches for the SM Higgs boson at the LHC~\cite{atlascombo,cmscombo} exclude a SM Higgs
boson with $M_H\leq 600\gev$, except for the narrow mass range $\approx 122-127\gev$.
In this mass range both the ATLAS and the CMS Collaborations observe
a significant excess of events in data at $M_H\approx 125\gev$ with local significances of 5.9 and 5.0 s.d., respectively. 
These excesses are formed by smaller excesses observed in searches focused on $\Hgg$ and $H\to ZZ^{(*)}$ decays, while no significant excesses 
have been found in searches targeting fermionic decay modes ($H\to b\bar{b}$ and $H\to \tau^+\tau^-$) with the datasets analyzed so far.

Searches for $\Hgg$ are particularly sensitive to new particles beyond the SM contributing to the loop-mediated $Hgg$ 
and/or $H\gamma\gamma$ vertices, and to deviations
in the couplings between the SM particles and the Higgs boson from those predicted by the SM. 
For example, alternative models of electroweak symmetry breaking~\cite{beyond-SM} can involve suppressed couplings of the
Higgs boson to fermions, with the extreme case being the fermiophobic Higgs boson ($H_{\rm f}$) scenario, 
in which $H_{\rm f}$ has no tree-level couplings to fermions but has SM coupling to weak gauge bosons. In this scenario the GF production 
mechanism is absent, decays into fermions are heavily suppressed, and ${\cal B}(\Hgg)$ is significantly enhanced.
The best-fit cross sections to the signal-like excesses in the $\Hgg$ searches at the LHC 
show small deviations of about 1.5 s.d. above the SM prediction~\cite{atlascombo,cmscombo}. A more detailed global fit to 
Higgs boson couplings~\cite{Plehn} shows no significant deviations. Hence, the analysis of more data is needed for more definitive conclusions.
Searches for a  fermiophobic Higgs boson were performed by the LEP Collaborations~\cite{LEP-FH}, the CDF~\cite{CDF-Hgg}
and D0~\cite{D0-Hgg} Collaborations and, most recently, by the ATLAS~\cite{ATLAS-Hgg-FH} and CMS~\cite{CMS-Hgg-FH} Collaborations.
The most restrictive limits result from the combination of $\Hgg$, $H\to W^+W^-$ and $H\to ZZ$ searches by the CMS Collaboration,
excluding the mass range $110<M_{H_{\rm f}}<194\gev$.

In this Article, we present the result from the search for a Higgs boson decaying into $\gamma\gamma$ using the complete 
dataset collected with the D0 detector in $p\bar{p}$ collisions at $\sqrt{s}=1.96\tev$ during Run II of the Tevatron
Collider.  This search employs multivariate techniques to improve the signal-to-background
discrimination, and is separately optimized for a SM Higgs boson and for a fermiophobic Higgs boson.
Compared to the previous D0 publication~\cite{D0-Hgg}, the sensitivity for the SM Higgs boson is improved by about 40\%,
resulting in the most restrictive limits to date from the Tevatron in this decay mode. The search for a fermiophobic Higgs boson
has comparable sensitivity with the most recent result from the CDF Collaboration~\cite{CDF-Hgg}. 
This result constitutes an important input for the upcoming publications on combinations 
of Higgs boson searches by the D0 experiment, as well as by both Tevatron experiments, using  the complete Run II dataset.

\section{D0 detector and data set}

The D0 detector is described in detail elsewhere~\cite{d0det}.
The subdetectors most relevant to this analysis are the central tracking system, 
composed of a silicon microstrip tracker (SMT) and a central fiber tracker (CFT) 
in a 2 T solenoidal magnetic field, the central preshower (CPS),
and the liquid-argon and uranium sampling calorimeter.

The SMT has about 800,000 individual strips, with typical pitch
of 50--80~$\mu$m, and a design optimized for tracking and
vertexing capability at pseudorapidities of $|\eta|<2.5$~\cite{d0_coordinate}.
The system has a six-barrel longitudinal structure, each with a set 
of four layers arranged axially around the beam pipe, and interspersed 
with 16 radial disks. In the summer of 2006 an additional layer of silicon sensors 
was inserted at a radial distance of $\approx 16$~mm from the beam axis, and 
the two outermost radial disks were removed.
The CFT has eight thin coaxial barrels, each 
supporting two doublets of overlapping scintillating fibers of 0.835~mm 
diameter, one doublet being parallel to the collision axis, and the 
other alternating by $\pm 3^{\circ}$ relative to the axis. Light signals 
are transferred via clear fibers to visible light photon counters (VLPC) 
that have about 80\% quantum efficiency.

The CPS is located just outside of the
superconducting magnet coil (in front of the calorimetry) and is formed by  
one radiation length of absorber followed by several layers of extruded 
triangular scintillator strips that are read out using wavelength-shifting 
fibers and VLPCs.

The calorimeter consists of three sections housed in separate cryostats: a central 
calorimeter covering up to $|\eta|\approx 1.1$, 
and two end calorimeters extending the coverage up to $|\eta|\approx 4.2$. 
Each section is divided into electromagnetic (EM) layers on the inside and hadronic layers on the outside.
The EM part of the calorimeter is segmented into four
longitudinal layers with transverse segmentation of $\Delta\eta \times \Delta\phi = 0.1\times 0.1$~\cite{d0_coordinate},
except in the third layer (EM3), where it is $0.05\times 0.05$. The calorimeter
is well suited for a precise measurement of electron and photon energies, 
providing a resolution of $\approx 3.6\%$ at energies of $\approx 50\gev$. 

Luminosity is measured using plastic scintillator arrays located in front 
of the end calorimeter cryostats, covering $2.7 < |\eta| < 4.4$. 
Trigger and data acquisition systems are designed to accommodate 
the high luminosities of Run II. Based on preliminary information from 
tracking, calorimetry, and muon systems, the output of the first level 
of the trigger is used to limit the rate for accepted events to about 
2~kHz. At the next trigger stage, with more refined 
information, the rate is reduced further to about 1~kHz. These
first two levels of triggering rely mainly on hardware and firmware.
The third and final level of the trigger, with access to all the event 
information, uses software algorithms and a computing farm, and reduces 
the output rate to about 100~Hz, which is written to tape.

This analysis uses the complete dataset of $p\bar{p}$ collisions at $\sqrt{s}=1.96\tev$ 
recorded with the D0 detector during Run II of the Tevatron Collider. 
The data are acquired using triggers requiring at least two clusters of energy 
in the EM calorimeter with loose shower shape requirements and varying transverse momentum ($\pt$) 
thresholds between $15\gev$ and $25\gev$. The trigger efficiency is close to 100\% for final 
states containing two photon candidates with $\pt > 25\gev$.
Only events for which all subdetector systems are fully operational are considered. 
The analyzed dataset corresponds to an integrated luminosity of $9.6$~fb$^{-1}$~\cite{d0lumi}. 

\section{Event simulation}
\label{sec:mc}

Monte Carlo (MC) samples of Higgs boson signal are generated separately for the
GF, VH and VBF processes using the {\sc pythia}~\cite{pythia} leading-order (LO) event generator 
with the CTEQ6L1~\cite{cteq} parton distribution functions (PDFs). Signal samples are
generated for $100\leq M_H\leq 150\gev$, in increments of $5\gev$.
Signal samples are normalized using the next-to-next-to-leading order (NNLO)
plus next-to-next-to-leading-logarithm (NNLL)  cross sections for GF~\cite{ggHxsect}
and NNLO for VH and VBF processes~\cite{VHxsect,VBFxsect}, computed with
the MSTW 2008 PDF set~\cite{signalPDF}. The Higgs boson's branching fraction predictions 
are from {\sc hdecay} \cite{hdecay}. To improve the signal modeling for the GF process, 
the $\pt$ of the Higgs boson is corrected to match the prediction at NNLO+NNLL
accuracy by the {\sc hqt} program~\cite{hqt}.
In the case of the fermiophobic model, where the GF process is absent, the VH and VBF cross sections 
are normalized to the SM prediction, while the modified $\Hgg$ branching fractions
are computed with {\sc hdecay}. 

The main background affecting this search is direct photon pair (DPP) production, 
where two isolated photons with high transverse momenta are produced.
The rest of the backgrounds are of instrumental origin and include $\gamma$+jet ($\gamma j$) and dijet ($jj$) production, where
at least one jet is misidentified as a photon. A smaller instrumental background originates
from $Z/\gamma^*\to e^+e^-$ production, where both electrons are misidentified as
photons. The normalization and shape of the $\gamma j$ and $jj$ backgrounds, as well
as the overall normalization of the DPP background, are estimated from data, as discussed
in Sect.~\ref{sec:background_modeling}. The shape of the DPP background is modeled via a
MC sample generated using {\sc sherpa}~\cite{sherpa} with the CTEQ6L1 PDF set.
Recent measurements of DPP differential cross sections~\cite{diphotonPLB} have shown that {\sc sherpa} 
provides an adequate model of this process in the kinematic region of interest for this search.
The $Z/\gamma^*\to e^+e^-$ process is modeled using {\sc alpgen}~\cite{alpgen} with the CTEQ6L1 PDF set,
interfaced to {\sc pythia} for parton showering and hadronization, with a subsequent correction
to the $\pt$ spectrum of the $Z$ boson to match measurements in data~\cite{Zqt}.
The $Z/\gamma^*\to e^+e^-$ MC sample is normalized to the NNLO theoretical cross section~\cite{Zxsec}.

All MC samples are processed through a {\sc geant}-based \cite{geant}
simulation of the D0 detector. To accurately model the effects 
of multiple $p\bar{p}$ interactions and detector noise, data events from random $p\bar{p}$ crossings
that have an instantaneous luminosity spectrum similar to the events in this analysis
are overlaid on the MC events. These MC events are then processed using the same reconstruction 
algorithms as used on the data. Simulated events are corrected so that the physics object identification 
efficiencies, energy scales and energy resolutions match those determined in data control samples.

\section{Object identification and event selection}

\subsection{Photon reconstruction and energy scale}
\label{sec:photonID}
Photon candidates are formed from clusters of calorimeter cells
within a cone of radius ${\cal R}=\sqrt{(\Delta\eta)^2+(\Delta\phi)^2}=0.4$ around a seed tower \cite{d0det}. 
The final cluster energy is then recalculated from the inner core with ${\cal R}=0.2$. 
The photon candidates are selected by requiring:
(i) at least $95\%$ of the cluster energy is deposited in the
EM calorimeter layers, (ii) the calorimeter isolation 
${\cal I} = [E_{\text{tot}}(0.4)-E_{\text{EM}}(0.2)]/E_{\text{EM}}(0.2)<0.1$,
where $E_{\text{tot}}(0.4)$ is the total energy in a cone of radius ${\cal R}=0.4$ and 
$E_{\text{EM}}(0.2)$ is the EM energy in a cone of radius ${\cal R}=0.2$, (iii) the scalar sum of 
the $\pt$  of all tracks ($p_{\rm T,trk}^{\rm sum}$) 
originating from the hard-scatter $p\bar{p}$ collision vertex (see Sect.~\ref{sec:pvID}) in an annulus of $0.05<{\cal R}<0.4$ around the EM cluster
is less than $2\gev$, and
(iv) the energy-weighted EM shower width
is required to be consistent with that expected for an electromagnetic shower.
This analysis only considers photon candidates with pseudorapidity $|\etag|<1.1$.

To suppress electrons misidentified as photons,
the EM clusters are required not to be spatially matched to significant tracker activity,
either a track, or a pattern of hits in the SMT and CFT consistent
with that of an electron or positron trajectory~\cite{HOR}. 
In the following, this requirement will be referred to as a ``track-match" veto.

\begin{figure}[t]
\centering
\includegraphics[width=0.45\textwidth]{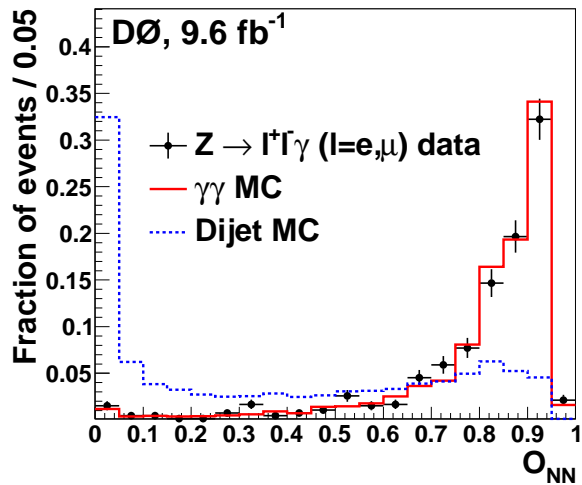}
\caption{Comparison of the normalized $\onn$ spectra for photons from DPP MC simulations and $Z\to \ell^+\ell^-\gamma$ data events (points with statistical error bars), and
for misidentified jets from simulated dijet events.}
\label{fig:photonANN}
\end{figure}

To suppress jets misidentified as photons, an artificial neural network (NN) discriminant, which 
exploits differences in tracker activity and energy deposits in the calorimeter and CPS
between photons and jets, is defined~\cite{ONN}.  The photon NN is trained using 
diphoton and  dijet MC samples generated using {\sc pythia}, using the following 
discriminating variables:  $p_{\rm T,trk}^{\rm sum}$,
the numbers of cells above a certain threshold requirement in the first EM calorimeter
layer within ${\cal R} < 0.2$ and within $0.2<{\cal R}<0.4$ of the EM cluster,
the number of associated CPS clusters within ${\cal R}< 0.1$ of
the EM cluster, and a measure of the width of the energy deposition in the CPS.
The performance of the photon NN  is verified  using a data event sample consisting 
of photons radiated from charged leptons in $Z$ boson decays 
($Z\to\ell^+\ell^-\gamma$, $\ell=e,\mu$)~\cite{Zg}.
Figure \ref{fig:photonANN} compares the NN output ($\onn$) distributions of photons and jets.
The shape of the $\onn$ distribution for photons is found to be in good agreement between
data and the MC simulation and is significantly different from the shape for misidentified
jets. The latter is validated using a sample enriched in jets misidentified as photons as
discussed in Sect.~\ref{sec:background_modeling}.
Photon candidates are required to have a $\onn$ value larger than 0.1, which
is close to 100\% efficient for photons while rejecting approximately 40\% of the remaining misidentified jets.

%%%%%%%%%%%%%%
%%%%%%%%%%%%%%

The measured photon energies are calibrated using a two-step correction procedure.
In the first step, the energy response of the calorimeter to photons is calibrated using
electrons from $Z$ boson decays. The resulting corrections are then applied to all 
electromagnetic clusters. Since electrons and photons shower
differently, with electrons suffering from a larger energy loss in material
upstream of the calorimeter, the application of this first set of corrections results in an overestimate
of the photon energy which depends on $\etag$. 
In the second step, additional corrections are derived for photons reconstructed in
the central calorimeter using a detailed {\sc geant}-based 
simulation  of the D0 detector response. These corrections are derived  as a function
of photon transverse momentum ($\ptg$) in seven intervals of $\etag$: $|\etag|<0.4$, $0.4\leq |\etag|<0.6$,
$0.6\leq |\etag|<0.7$, $0.7\leq |\etag|<0.8$, $0.8\leq |\etag|<0.9$, $0.9\leq |\etag|<1.0$, and $1.0\leq |\etag|<1.1$, and
separately for photons with and without a matched CPS cluster.  
The per-photon probability to have a matched CPS cluster is measured using photons radiated from  charged 
leptons in $Z$ boson decays  ($Z \to \ell^+\ell^-\gamma$, $\ell=e,\mu$) and is $\approx 73\%$.
The finer binning at higher $\eta$ is motivated by the strong dependence of the
energy-loss corrections for electrons on $\eta$.
The resulting corrections for photons with (without) a matched CPS cluster are 
largest at low $\ptg$ $\approx 20\gev$ and range from about $-1.5\%$ 
in the $|\etag|<0.4$ interval, to about $-6\%$ ($-10\%$) in the $|\etag|\geq 1.0$ interval.

\subsection{Primary vertex reconstruction}
\label{sec:pvID}

At the Tevatron the distribution of $p\bar{p}$ collision vertices has a Gaussian width of about 25~cm.
The proper reconstruction of the event kinematics, in particular $\ptg$ and thus $\mgg$, requires the reconstruction and then correct selection of 
the hard-scatter $p\bar{p}$ collision primary vertex (PV) among the various candidate
PVs originating from additional $p\bar{p}$ interactions. 

The algorithm used for PV reconstruction is described 
in detail elsewhere~\cite{bidnim}. In a first step, tracks with two or more associated SMT hits and $\pt > 0.5\gev$ 
are clustered along the $z$ direction. This is followed by a Kalman Filter fit~\cite{kalman} to a common vertex
of the tracks in each of the different vertices. Events are required to have at least one reconstructed PV with
a $z$ coordinate ($z_{\rm PV}$) within 60 cm from the center of detector, a requirement that is $\approx 98\%$ efficient.

The selection of the hard-scatter PV from the list of PV candidates with $|z_{\rm PV}|<60$~cm 
is based on an algorithm exploiting both the track multiplicity of
the different vertices and the transverse and longitudinal
energy distributions in the EM calorimeter and the CPS. 
These energy distributions allow the estimation of the photon direction and thus 
the $z$ coordinate of its production vertex along the beam direction.
When one or both photons reconstructed in the EM
calorimeter also deposit part of their energy in the
CPS, the algorithm chooses the PV whose $z_{\rm PV}$ is
closest to the extrapolation of the
photon trajectory determined from the calorimeter and
the CPS information~\cite{pointing}, provided the distance between
the coordinates of the vertex and of the photon
trajectory is smaller than 3 s.d.
The uncertainty on this distance is dominated by
the uncertainty on the extrapolation of the photon
direction, which ranges from $\approx 2.5$~cm for
photons with $|\etag|<0.4$ to $\approx 4.3$~cm for 
photons with $|\etag|>0.8$. Otherwise, the algorithm 
chooses the PV with the largest
multiplicity of associated tracks.

\begin{figure}[t]
\centering
\includegraphics[width=0.45\textwidth]{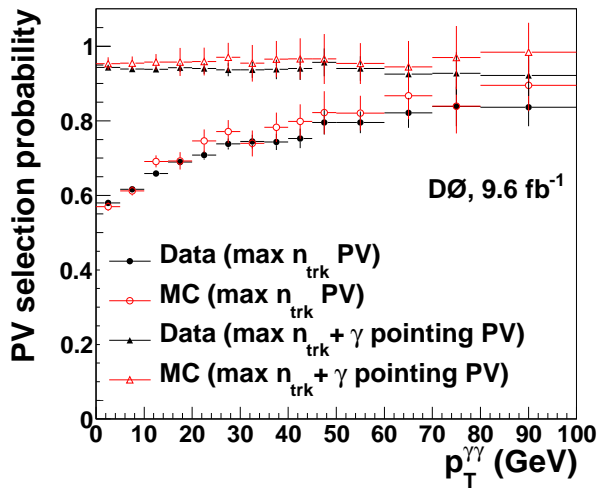}
\caption{Probability to select the correct hard-scatter PV as a function of $\ptgg$ as measured
in $Z/\gamma^{*}\to e^+e^-$ events excluding the electron and positron tracks from consideration. The two 
different algorithms discussed in the text are compared.}
\label{fig:pvalgo}
\end{figure}

This algorithm is optimized using $Z/\gamma^{*}\to e^+e^-$ data events, 
where the correct hard-scatter PV associated with the
reconstructed tracks is treated as corresponding to a diphoton
event by ignoring the track information from the $e^+e^-$ pair,
and added to the list of PV candidates to which the
selection algorithm will be applied. The fraction of $Z/\gamma^{*}\to e^+e^-$ events for which
the selected PV agrees with the known hard-scatter PV is shown in Fig.~\ref{fig:pvalgo} 
as a function of diphoton transverse momentum ($\ptgg$) for two different hard-scatter PV
selection algorithms. For an algorithm selecting the hard-scatter PV as the one with the highest track multiplicity,
the average selection probability is only
$\approx 65\%$ and shows a significant dependence on $\ptgg$.
The improved algorithm used in this analysis, including also
photon pointing information, achieves an average selection probability
of $\approx 95\%$, almost constant as a function of $\ptgg$.

%%%%%%%%%%%%%%
%%%%%%%%%%%%%%

\subsection{Event selection}

At least two photon candidates satisfying the requirements listed in Sect.~\ref{sec:photonID} and having $\ptg>25\gev$ and $|\etag|<1.1$ are required. 
If more than two photon candidates are identified, only the two photon candidates with highest $\ptg$ are considered.
At least one of the photon candidates in each event is required to have a matched CPS cluster. 
The photon kinematic variables are computed with respect to the vertex selected using the algorithm described in Sect.~\ref{sec:pvID}.
A requirement of $\mgg>60\gev$ is made to ensure a trigger efficiency close to 100\%.

The acceptance of the kinematic requirements is $\approx 42\%$, as estimated by applying the
$\ptg$ and $\etag$ requirements to generated photons in a $gg\to \Hgg$  MC sample assuming $M_H=125\gev$.
At the same assumed $M_H$, the overall event selection efficiency, taking into account acceptance and
reconstruction, identification and selection efficiencies, is $\approx 22\%$, almost independent on the
signal production mechanism.

To improve the sensitivity to signal, events are categorized into two statistically independent samples with
different signal-to-background ratios. Events where both photon candidates satisfy $\onn>0.75$ (``photon-enriched" sample)
and events where at least one photon candidate satisfies  $0.1<\onn<0.75$  (``jet-enriched" sample) are analyzed separately.
The corresponding sample compositions are discussed in Sect.~\ref{sec:background_modeling}.

\subsection{Invariant mass reconstruction}
\label{sec:invmass}

\begin{figure}[t]
\centering
\includegraphics[width=0.45\textwidth]{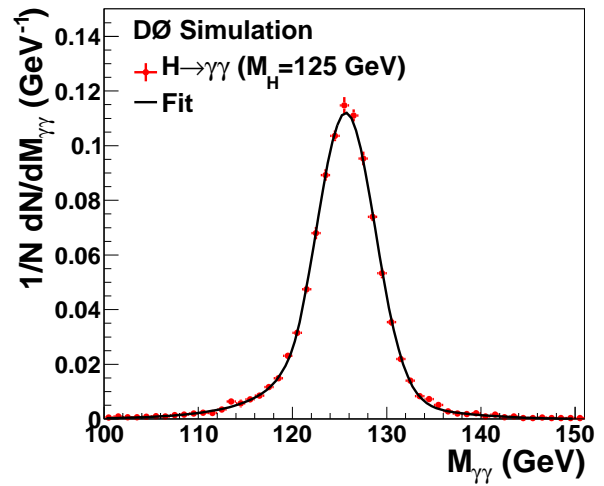}
\caption{Distribution of the reconstructed diphoton invariant mass distribution corresponding to a Higgs boson signal with $M_H=125\gev$.
  The line shows the result of a fit to the distribution using the functional form described in Sect.~\ref{sec:invmass}.}
\label{fig:mgg_signal}
\end{figure}

After the selection of the $p\bar{p}$ collision vertex and the photon energy scale corrections, the $\mgg$ 
distribution for a Higgs boson signal follows a Gaussian distribution peaking at the generated Higgs boson mass, with 
small non-Gaussian tails. This distribution can be modeled by the sum of a Crystal Ball function~\cite{CrystalBall}, 
describing a narrow Gaussian core and a power-law tail toward lower masses,  and a wider Gaussian distribution, 
describing tails from misvertexing or imperfect photon energy scale corrections. Figure~\ref{fig:mgg_signal} shows such 
a fit to the inclusive $\mgg$ spectrum for signal MC with $M_H=125\gev$.
The resolution of the Gaussian core is found to be $\approx 3.1\gev$, and varies by $\pm 13\%$ when varying
$M_H$ by $\pm 25\gev$.

%%%%%%%%%%%%%%
%%%%%%%%%%%%%%

\section{Background modeling and sample composition}
\label{sec:background_modeling}
 
The normalization and shape of the $Z/\gamma^*\to e^+e^-$ background are estimated using 
simulation. Electrons are misidentified as photons at a rate of about 2\%
due to track reconstruction inefficiencies. Such tracking inefficiency is measured
in data using a ``tag-and-probe" method, where $Z\to e^+e^-$  events
are selected with one of the electrons (``tag") passing all identification criteria,
including matching of the track to the calorimeter cluster, while only calorimeter requirements are applied to 
the other electron (``probe"). The electron misidentification rate is computed
as the fraction of events where the probe electron satisfies the ``track-match" veto requirement defined in Sect.~\ref{sec:photonID}.
The misidentification rate measured in data in this way is applied to the simulated $Z/\gamma^*\to e^+e^-$ sample.
 
The $\gamma j$ and $jj$ yields are estimated using a data-driven method~\cite{bkg-subtract} (``matrix method").
For selected events, the two photons are separated into two types: those with $\onn>0.75$ (well-identified photon, ``p") and those with $0.1<\onn<0.75$ (likely fake photon, ``f").
Events are then classified in four categories:  (i) two type-p photons, (ii) the higher $\ptg$ (leading) photon is type p and the lower $\ptg$ (trailing) photon is type f, 
(iii)  the leading photon is type f and the trailing photon is type p, and (iv) two type-f photons.
The corresponding numbers of events, after subtracting the $Z/\gamma^*\to e^+e^-$ contribution, 
are denoted as $N_{\rm pp}$, $N_{\rm pf}$, $N_{\rm fp}$ and $N_{\rm ff}$.
The different efficiencies of the $\onn>0.75$ requirement for photons ($\epsilon_{\gamma}$) 
and jets ($\epsilon_j$) are used to estimate the sample composition by solving a system of linear equations:
\begin{equation}
\label{matrix1}
(N_{\gamma \gamma}, N_{\gamma j}, N_{j \gamma}, N_{jj})  = (N_{\rm pp}, N_{\rm pf}, N_{\rm fp}, N_{\rm ff}) \times {\boldsymbol{{\cal E}}}^{-1},
\end{equation}
where $N_{\gamma\gamma}$ ($N_{jj}$) is the number of $\gamma\gamma$ ($jj$) events and $N_{\gamma j}$ ($N_{j \gamma}$)
is the number of $\gamma j$ events with the leading (trailing) cluster as the photon.
The $4\times4$ matrix $\boldsymbol{{\cal E}}$ is constructed with the efficiency terms
$\epsilon_{\gamma}$ and $\epsilon_j$, parameterized as a function of $|\etag|$ for each photon candidate as 
determined from photon and jet MC samples, respectively. The $\epsilon_{\gamma}$ and $\epsilon_j$ efficiencies averaged over $|\etag|$ are
$\approx 76\%$ and $\approx 35\%$, respectively.
The efficiency $\epsilon_{\gamma}$ is validated with a data sample of photons radiated from 
charged leptons in $Z$ boson decays ($Z \to \ell^+\ell^-\gamma$, $\ell=e,\mu$).
The efficiency $\epsilon_j$ is validated using two independent control data samples
enriched in jets misidentified as photons, either by inverting the photon isolation variable (${\cal I}>0.1$), or
by requiring at least one track in a cone of ${\cal R}<0.05$ around the photon~\cite{diphoton-Xsection}.
In the following, the sum of  $\gamma j$ and $j\gamma$ contributions will be denoted as $\gamma j$ for simplicity.
The shapes of kinematic distributions for $\gamma j$ ($jj$) background are obtained from 
independent control samples by requiring one (two) photon candidate(s) to satisfy $\onn<0.1$.
The $\onn<0.1$ requirement leads to a mis-modeling of the $\etag$ spectrum, due to 
the  $|\etag|$ dependence of $\epsilon_j$. This is corrected by assigning a weight factor defined as $\epsilon_j(|\etag|)$/(1-$\epsilon_j(|\etag|)$) 
for each of the photon candidates with $\onn<0.1$. 

\begin{figure}[t]
\centering
\includegraphics[width=0.45\textwidth]{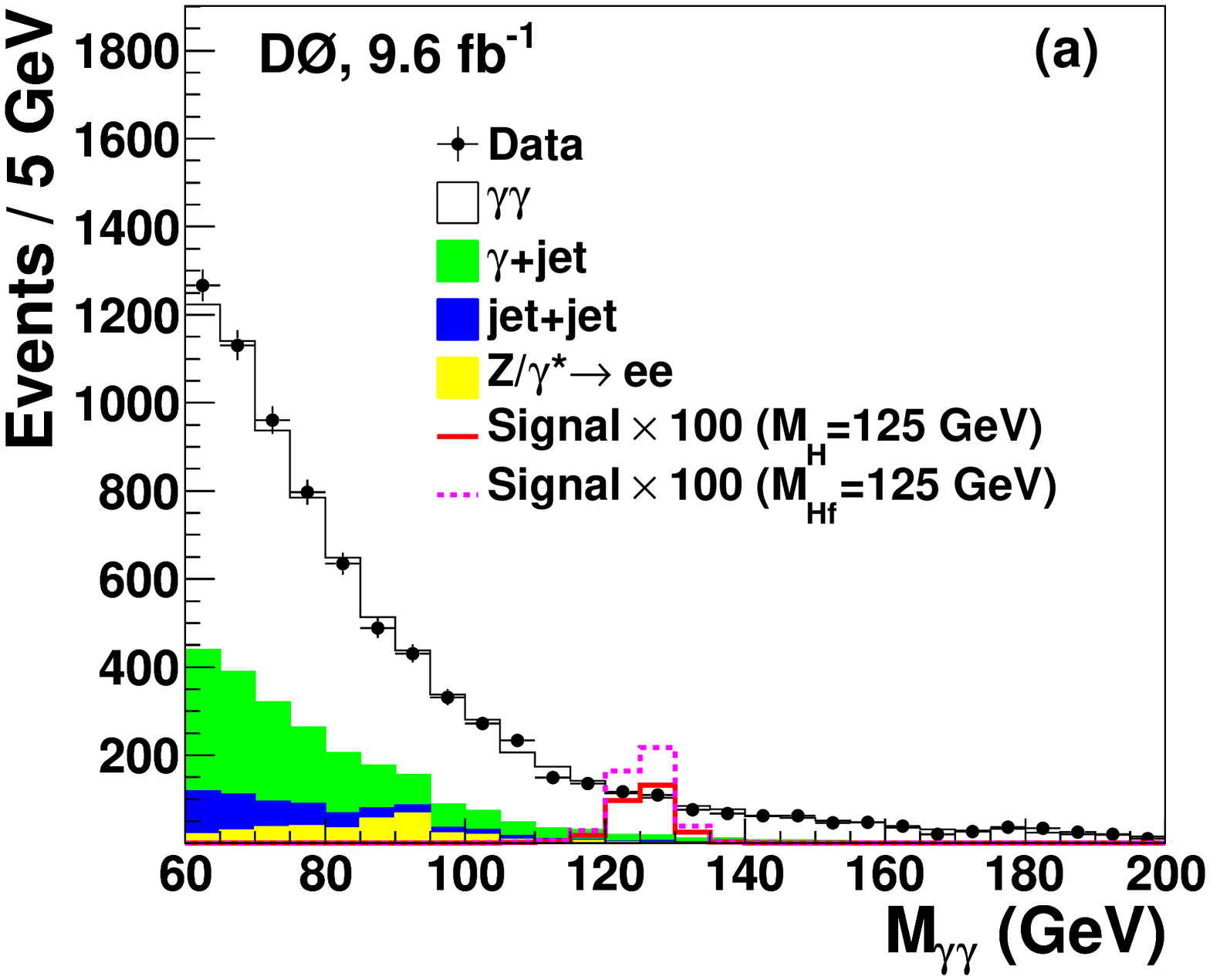}
\includegraphics[width=0.45\textwidth]{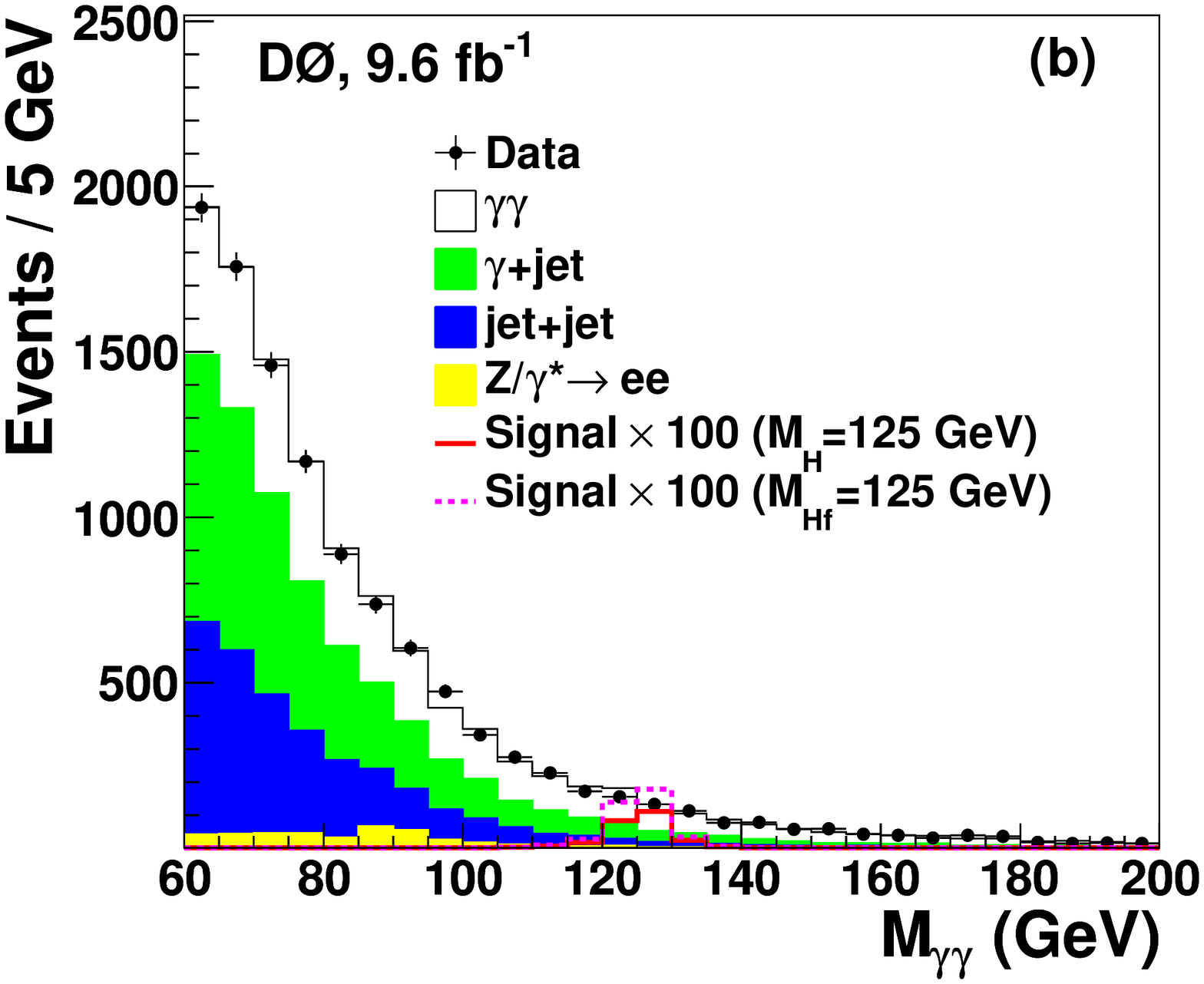}
\caption{\small (color online). Distribution of $\mgg$ in (a) the photon-enriched sample and (b) the jet-enriched sample.
The data (points with statistical error bars) are compared to the background prediction, broken down into its individual components.
The expected distributions for a SM Higgs boson and a fermiophobic Higgs boson with $M_H=125\gev$  
are also shown scaled by a factor of 100.}
\label{fig:mgg}
\end{figure}

As discussed in Sect.~\ref{sec:mc}, the kinematics of the DPP background are predicted using {\sc sherpa}. 
Since the estimated $N_{\gamma\gamma}$ from solving Eq.~\ref{matrix1} could 
include a contribution from signal events, it is only used as a prior normalization for the DPP background
to compare between data and background prediction.
The normalization of the DPP background is ultimately determined from an unconstrained fit to
the final discriminants used for hypothesis testing in both the photon-enriched and jet-enriched samples.
For each of these samples, two distributions are considered: a multivariate discriminant (see Sect.~\ref{sec:mva}) constructed to maximize
the separation between signal and background for events with $\mgg$ falling in the interval $M_H \pm 30\gev$ ("search region"), 
and the $\mgg$ spectrum for events outside this interval ("sideband region") that provide a high-statistics background-dominated
sample.
A comparison between data and the background prediction for the $\mgg$ spectrum, separately in the
photon-enriched and the jet-enriched samples, is shown in Fig.~\ref{fig:mgg}.

%%%%%%%%%%%%%%
%%%%%%%%%%%%%%

Tables~\ref{tab:yields_photonsample} and~\ref{tab:yields_jetsample} summarize the number of data events,
expected backgrounds, and expected SM and fermiophobic Higgs boson signals, resulting from the fit for five hypothesized Higgs boson masses, 
for the photon-enriched and jet-enriched samples, respectively.
For $M_H=125\gev$, the estimated background composition for the photon-enriched sample in the $\mgg$ interval of [$95\gev$, $155\gev$]
is about 80\% (DPP), 14\% ($\gamma j$), 3\% ($jj$) and 3\% ($Z/\gamma^*\to e^+e^-$).
The corresponding composition for the jet-enriched sample is about 48\% (DPP), 31\% ($\gamma j$), 18\% ($jj$) and 3\% ($Z/\gamma^*\to e^+e^-$).

%%%%%%%%
\begin{table*}[htbp]
\centering
\begin{tabular}{lccccc}
\hline \hline
$M_H$ (\gev) &105 & 115 & 125 & 135 & 145  \\ \hline
$\gamma\gamma$ (DPP) & 2777 $\pm$ 65 & 1928 $\pm$ 44 & 1355 $\pm$ 31 & 980 $\pm$ 22  & 721 $\pm$ 17 \\
$\gamma j$       & 704  $\pm$ 40 &  407 $\pm$ 24 &  238 $\pm$ 14 & 144 $\pm$ 9   &  88 $\pm$ 6 \\
$jj$           & 183  $\pm$ 16 &   93 $\pm$ 9  &   54 $\pm$ 6  &  34 $\pm$ 4   &  19 $\pm$ 2 \\
$Z/\gamma^{*} \to e^+e^-$  & 219  $\pm$ 40 &  149 $\pm$ 30 &   51 $\pm$ 11 &  22 $\pm$ 5   &  11 $\pm$ 3 \\
\hline
Total background     & 3883  $\pm$ 61 & 2577 $\pm$ 45 & 1698 $\pm$ 30 & 1180 $\pm$ 21 & 839 $\pm$ 16 \\
Data                 & 3777  & 2475 & 1664 & 1147& 813 \\
\hline
$H$ signal           & 3.6 $\pm$ 0.4 & 3.5 $\pm$ 0.4 & 3.0 $\pm$ 0.4 & 2.2 $\pm$ 0.3 & 1.4 $\pm$ 0.2 \\
$H_{\rm f}$ signal       & 49.8 $\pm$ 1.1 & 14.0 $\pm$ 0.3 & 4.8 $\pm$ 0.1 & 1.9 $\pm$ 0.1 & 0.79 $\pm$ 0.03 \\
\hline \hline
\end{tabular}
\caption{\label{tab:yields_photonsample} \small Signal, backgrounds and data yields for the photon-enriched sample within the $M_H \pm 30\gev$ mass window,
for $M_H=105\gev$ to $M_H=145\gev$ in $10\gev$ intervals. The background 
yields are from a fit to the data. The uncertainties include both statistical and systematic contributions added in quadrature and take into
account correlations among processes. The uncertainty on the total background is smaller than the sum in quadrature
of the uncertainties in the individual background sources due to the anti-correlation resulting from the fit.}
\end{table*}
%%%%%%%%

%%%%%%%%
\begin{table*}[htbp]
\centering
\begin{tabular}{lccccc}
\hline \hline
$M_H$ (\gev) &105 & 115 & 125 & 135 & 145  \\ \hline
$\gamma\gamma$ (DPP) & 1969 $\pm$ 47  & 1406 $\pm$ 33 & 1012 $\pm$ 24 & 734 $\pm$ 17 & 545 $\pm$ 13\\
$\gamma j$       & 1852 $\pm$ 100 & 1101 $\pm$ 60 &  653 $\pm$ 36 & 391 $\pm$ 22 & 251 $\pm$ 15\\
$jj$           & 1188 $\pm$ 94  &  647 $\pm$ 54 &  365 $\pm$ 31 & 219 $\pm$ 19 & 135 $\pm$ 12\\
$Z/\gamma^{*} \to  e^+e^-$  & 227 $\pm$ 39  &  152 $\pm$ 28 &   61 $\pm$ 11 &  30 $\pm$ 7  &  20 $\pm$  5\\
\hline
Total background     & 5236 $\pm$ 67 & 3307 $\pm$ 45 & 2091 $\pm$ 29 & 1374 $\pm$ 21 & 951 $\pm$ 17\\
Data                & 5287  & 3384 & 2156 & 1422 & 989\\
\hline
$H$ signal          & 2.7 $\pm$ 0.3 & 2.6 $\pm$ 0.3 & 2.2 $\pm$ 0.3 & 1.7 $\pm$ 0.2 & 1.1 $\pm$ 0.1\\
$H_{\rm f}$ signal      & 34.8 $\pm$ 0.8 & 9.8 $\pm$ 0.3 & 3.4 $\pm$ 0.1 & 1.34 $\pm$ 0.04 & 0.56 $\pm$ 0.02\\
\hline \hline
\end{tabular}
\caption{\label{tab:yields_jetsample} \small Signal, backgrounds and data yields for the jet-enriched sample within the $M_H \pm 30\gev$ mass window,
for $M_H=105\gev$ to $M_H=145\gev$ in $10\gev$ intervals. The background 
yields are from a fit to the data. The uncertainties include both statistical and systematic contributions added in quadrature and take into
account correlations among processes. The uncertainty on the total background is smaller than the sum in quadrature
of the uncertainties in the individual background sources due to the anti-correlation resulting from the fit.}
\end{table*}
%%%%%%%%

\section{Signal-to-background discrimination}
\label{sec:mva}

\begin{figure}[t]
\centering
\includegraphics[width=0.45\textwidth]{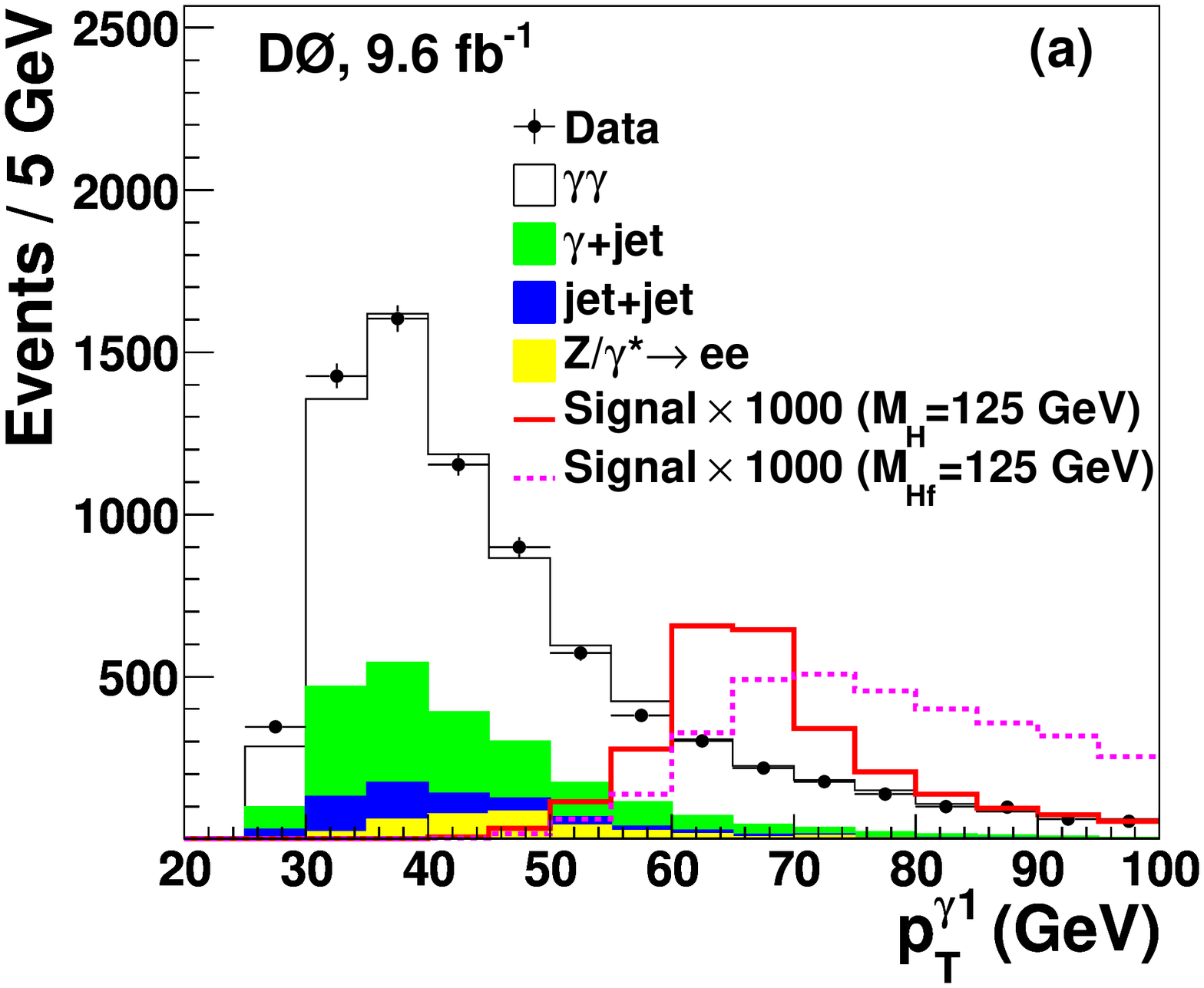}
\includegraphics[width=0.45\textwidth]{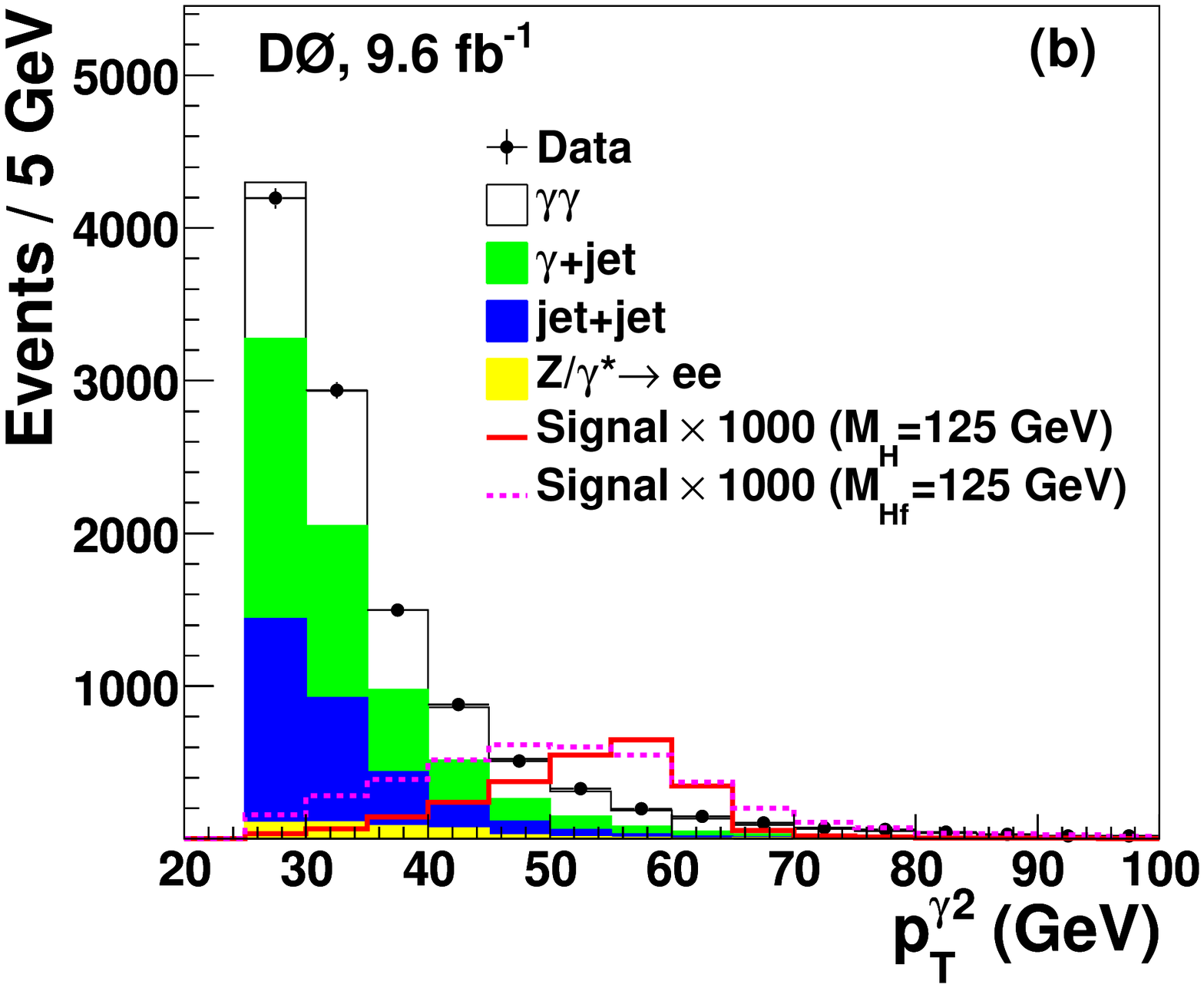}
\caption{\small (color online). Distribution of (a) $\pt^{\gamma 1}$ in the photon-enriched sample and (b) $\pt^{\gamma 2}$ in the jet-enriched sample.
The data (points with statistical error bars) are compared to the background prediction, broken down into its individual components.
The expected distributions for a SM Higgs boson and a fermiophobic Higgs boson with $M_H=125\gev$  
are also shown scaled by a factor of 1000.
These two BDT input variables are used in both the photon-enriched and jet-enriched samples, but are displayed here 
for only one of the samples for illustrative purposes.}
\label{fig:ptg}
\end{figure}

\begin{figure}[t]
\centering
\includegraphics[width=0.45\textwidth]{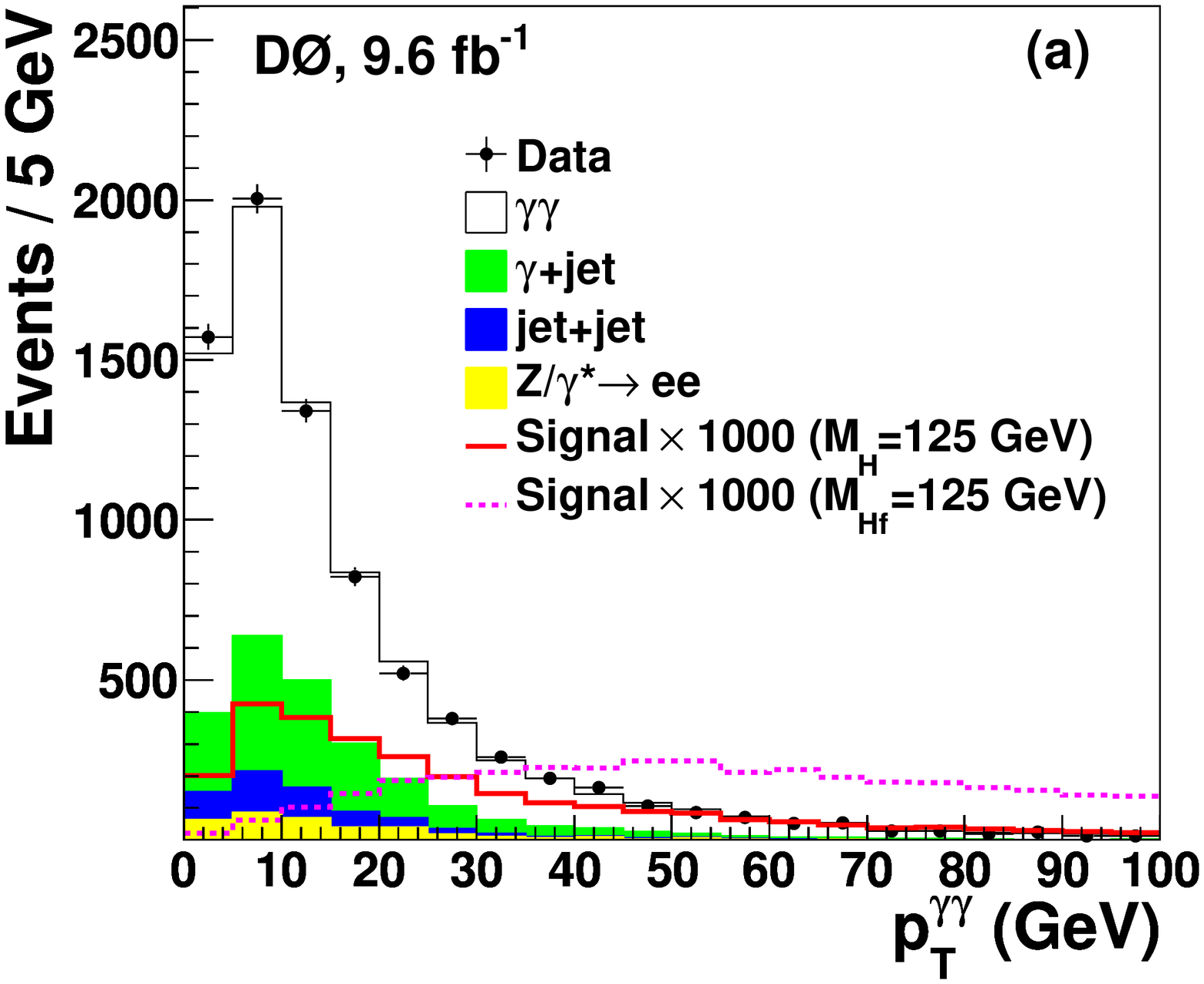}
\includegraphics[width=0.45\textwidth]{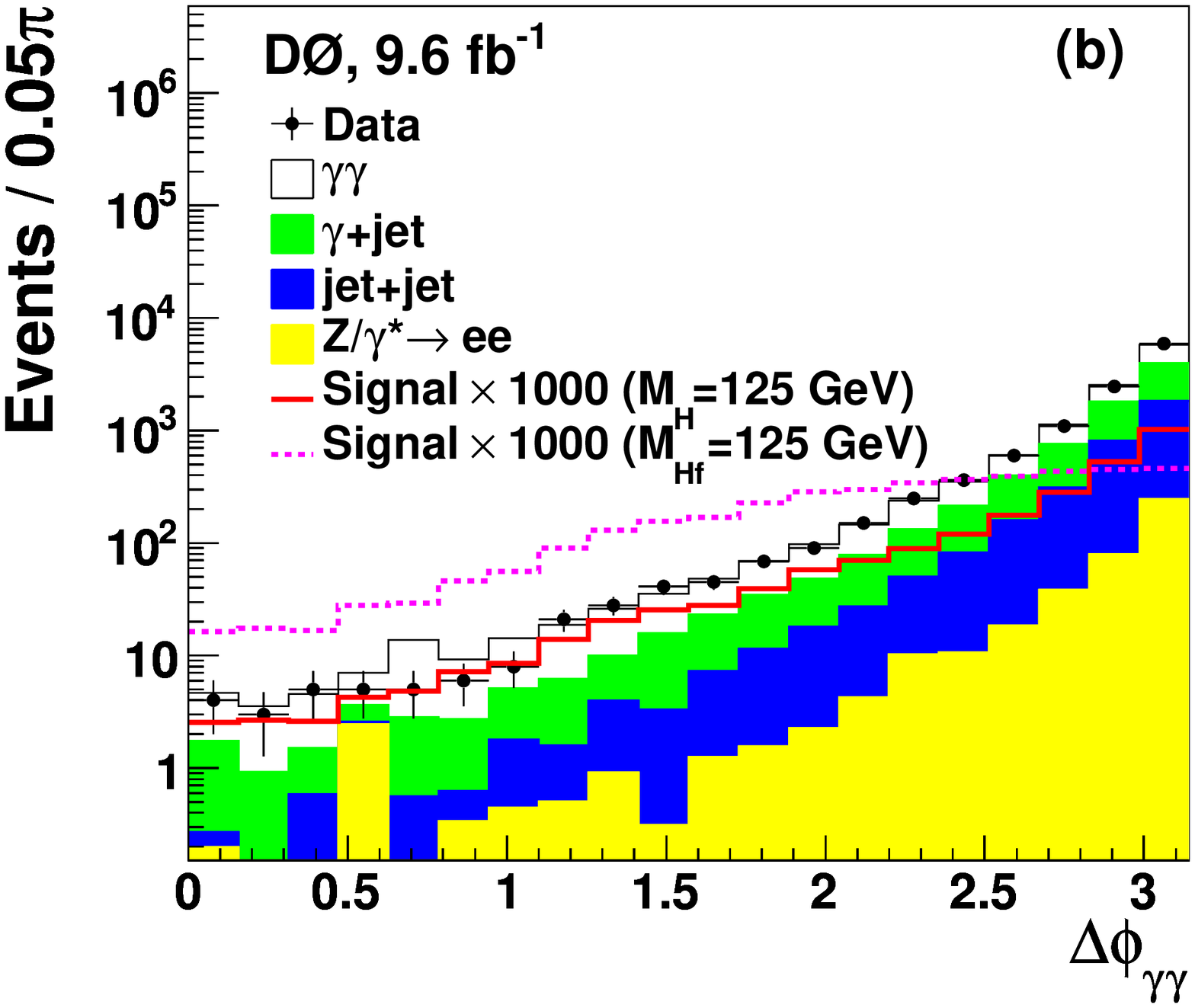}
\caption{\small (color online). Distribution of (a) $\ptgg$ in the photon-enriched sample and (b) $\dphigg$ in the jet-enriched sample.
The data (points with statistical error bars) are compared to the background prediction, broken down into its individual components.
The expected distributions for a SM Higgs boson and a fermiophobic Higgs boson with $M_H=125\gev$  
are also shown scaled by a factor of 1000.
These two BDT input variables are used in both the photon-enriched and jet-enriched samples, but are displayed here 
for only one of the samples for illustrative purposes.}
\label{fig:ptgg}
\end{figure}

The diphoton mass $\mgg$ is the most effective discriminating variable between the Higgs boson
signal and the background. However, further discrimination can be achieved
by exploiting additional kinematic variables as well as photon quality variables. A total
of ten well-modeled discriminating variables are considered in this search.
Two of these variables correspond to kinematic properties of the photons:
leading photon transverse momentum ($\pt^{\gamma 1}$) and trailing photon transverse 
momentum ($\pt^{\gamma 2}$) which, as illustrated in Fig.~\ref{fig:ptg},  follow a
harder spectrum in signal than in background, as expected for the decay of a 
heavy resonance.
Three of the variables are related to the kinematics of the diphoton system:
$\mgg$, $\ptgg$ and azimuthal angle
separation between the photons ($\dphigg$). The two latter variables give discrimination 
due to the large $\pt$ of the Higgs boson in VH and VBF production. Therefore,
as illustrated in Fig.~\ref{fig:ptgg}, $\ptgg$ and $\dphigg$ are particularly sensitive 
variables in the search for a fermiophobic Higgs boson.

\begin{figure}[t]
\centering
\includegraphics[width=0.45\textwidth]{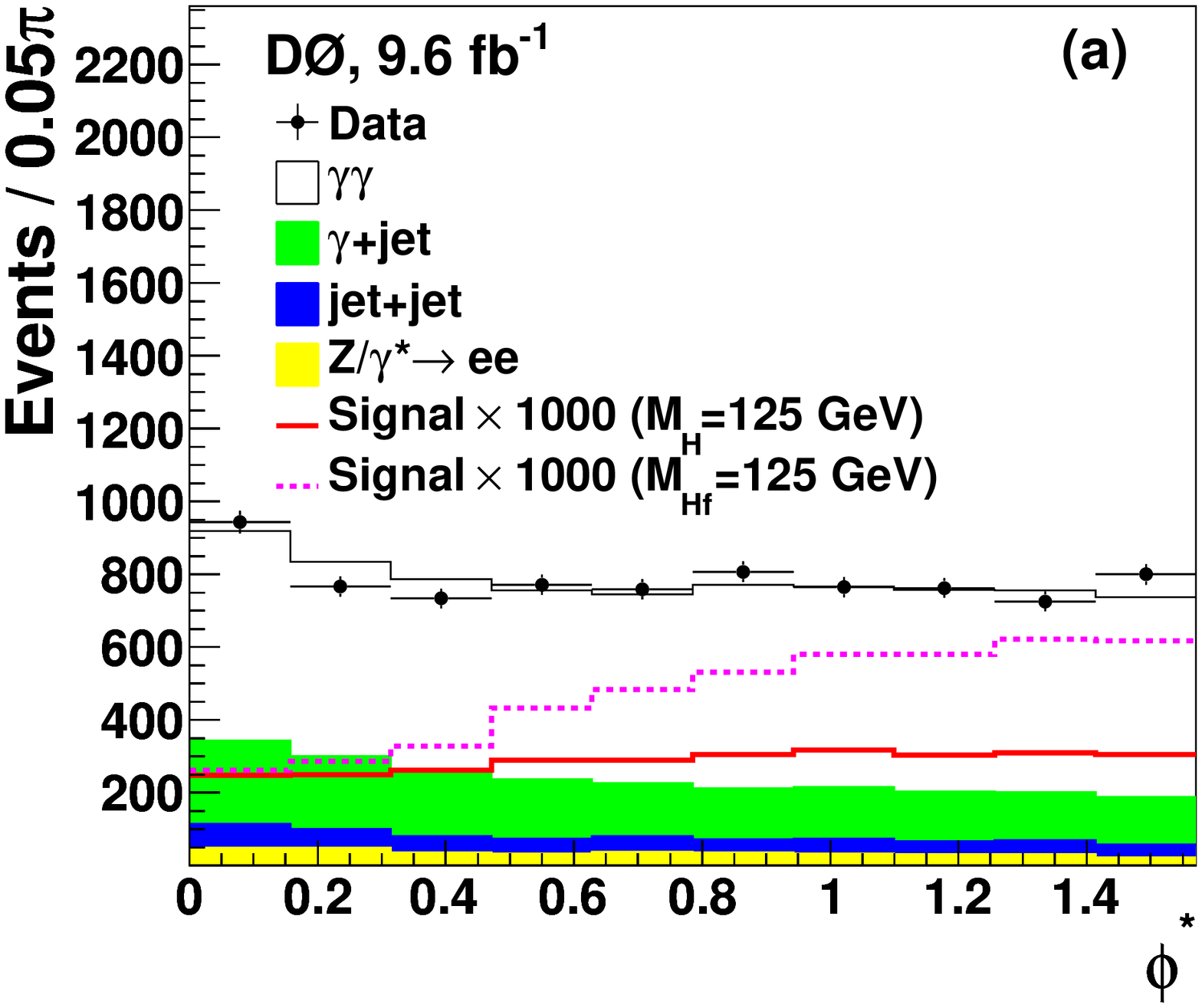}
\includegraphics[width=0.45\textwidth]{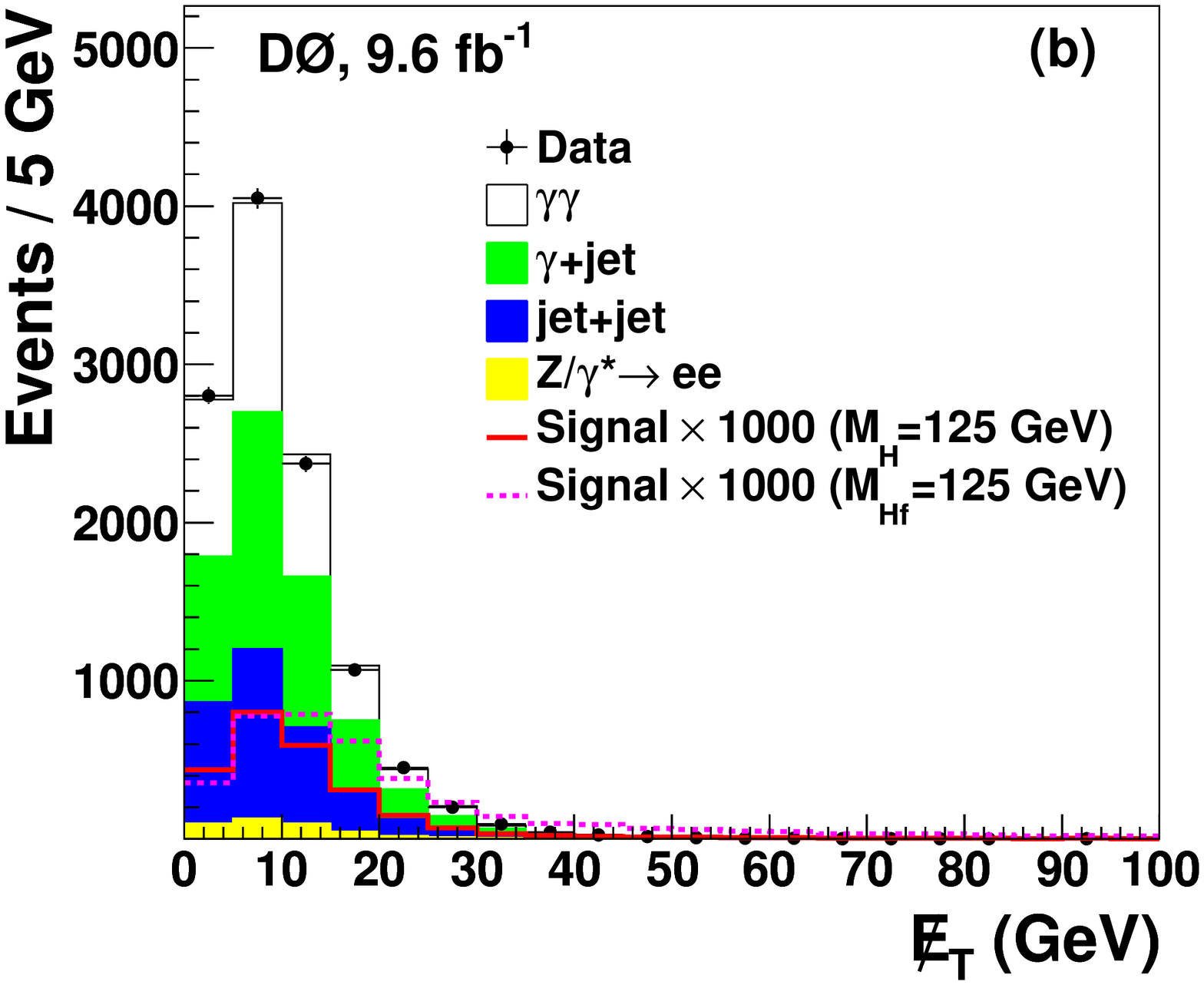}
\caption{\small (color online). Distribution of (a) $\phi^{*}$ in the photon-enriched sample and (b) $\met$ in the jet-enriched sample.
The data (points with statistical error bars) are compared to the background prediction, broken down into its individual components.
The expected distributions for a SM Higgs boson and a fermiophobic Higgs boson with $M_H=125\gev$  
are also shown scaled by a factor of 1000.
These two BDT input variables are used in both the photon-enriched and jet-enriched samples, but are displayed here 
for only one of the samples for illustrative purposes.}
\label{fig:phistar}
\end{figure}

The scalar nature of the Higgs boson affects the angular distributions
of the photons in the diphoton rest frame. To minimize uncertainties from
the transverse momentum of the colliding partons, the Collins-Soper 
frame~\cite{collins-soper-frame} is used. In this frame, the $z$ axis is defined 
as the bisector of the proton beam momentum and the negative of the antiproton 
beam momentum when they are boosted into the center-of-mass frame 
of the diphoton pair. The variable $\theta^{*}$ is defined as the angle between the leading 
photon momentum and the $z$ axis. The variable $\phi^{*}$ is defined as the angle 
between the diphoton plane and the $p\bar{p}$ plane. Due to the restriction to photons with $|\etag|<1.1$ in
this analysis, the $\cos\theta^{*}$ distribution has little discrimination between signal
and background, although it is considered in the search. In contrast, the angle $\phi^{*}$
provides useful discrimination between signal and background, particularly
for a fermiophobic Higgs boson, as illustrated in Fig.~\ref{fig:phistar}(a).

A significant fraction of $W$ and $Z$ boson decays in {\sl VH} production involves 
neutrinos that result in large missing transverse  energy ($\met$) in the final state. 
In contrast,  the $\met$ in background events is typically low, and mostly resulting
from jet energy mismeasurements. The $\met$ distribution 
in the jet-enriched sample is shown in Figure~\ref{fig:phistar}(b).
The $\met$ is reconstructed as the negative of the vectorial sum of the 
$\pt$ of calorimeter cells, and is corrected for the $\pt$ of identified muons
and the energy corrections to reconstructed jets in the calorimeter~\cite{jets}.

%%%%%%%%%%%%%%
%%%%%%%%%%%%%%

%%%%%%%%%%%%%%
%%%%%%%%%%%%%%

%%%%%%%%%%%%%%
%%%%%%%%%%%%%%

%%%%%%%%%%%%%%
\begin{figure}[t]
\centering
\includegraphics[width=0.45\textwidth]{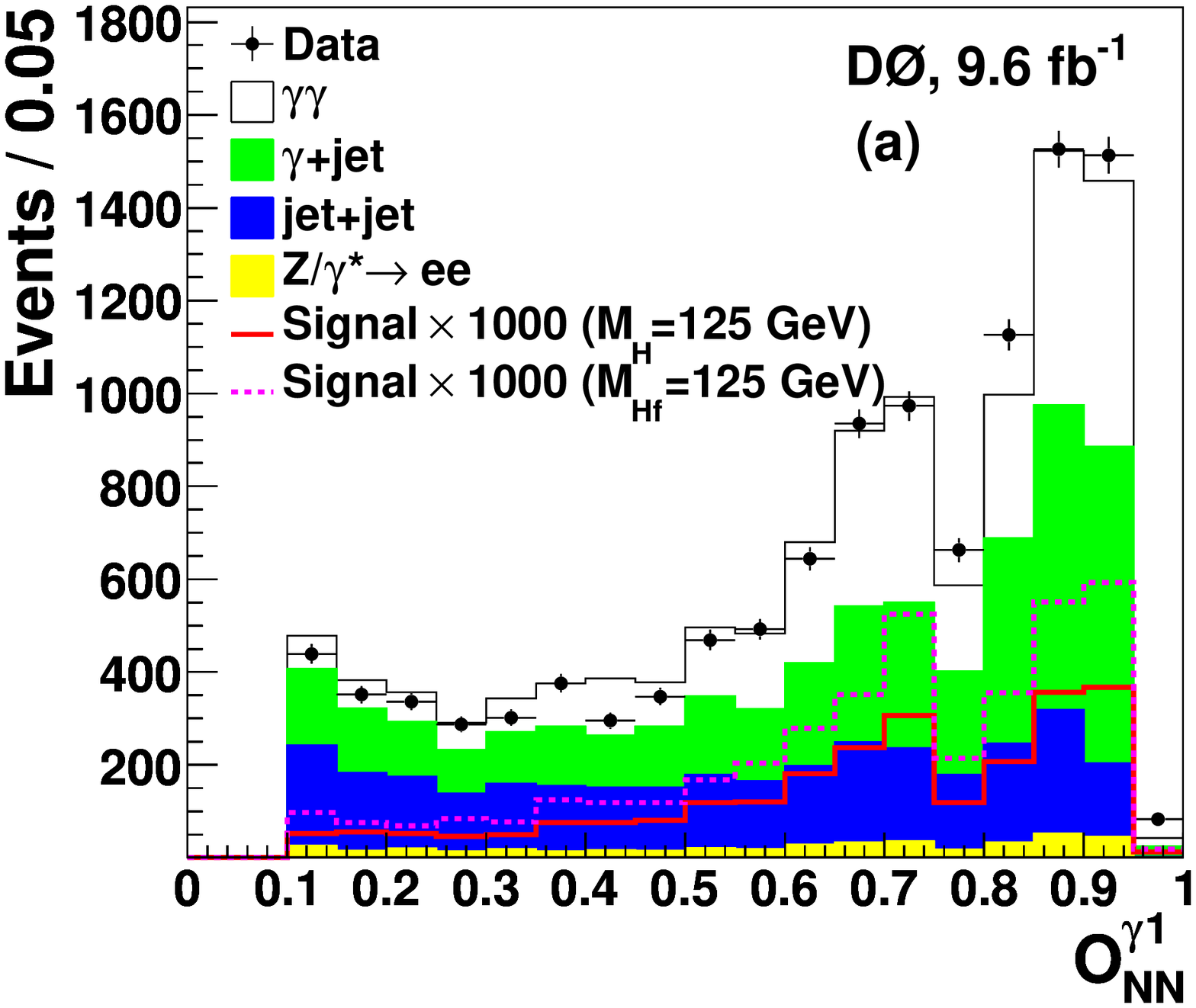}
\includegraphics[width=0.45\textwidth]{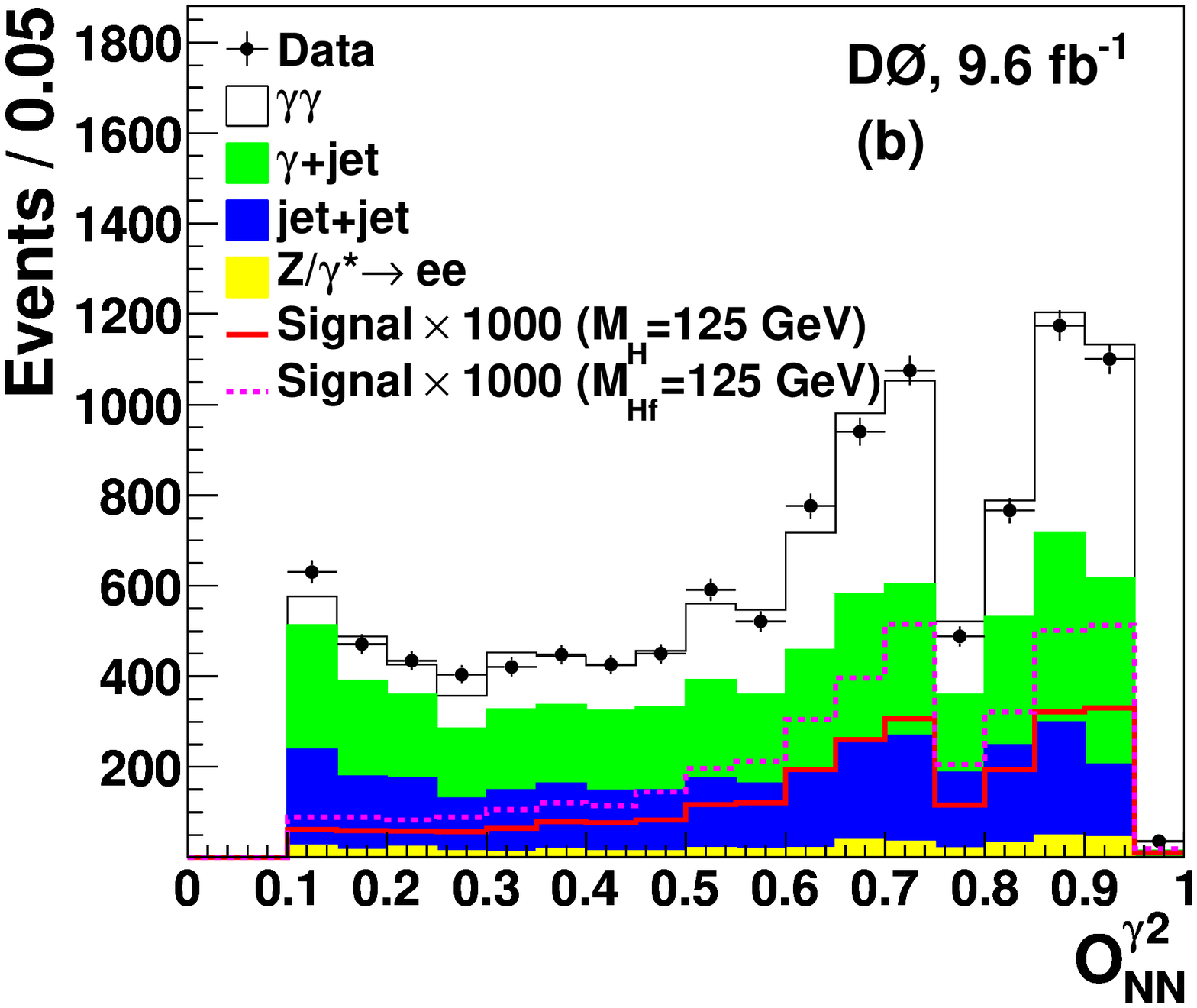}
\caption{\small (color online). Distribution of (a) $O_{\rm NN}^{\gamma 1}$ and (b) $O_{\rm NN}^{\gamma 2}$ in the jet-enriched sample.
The data (points with statistical error bars) are compared to the background prediction, broken down into its individual components.
The expected distributions for a SM Higgs boson and a fermiophobic Higgs boson with $M_H=125\gev$  
are also shown scaled by a factor of 1000.
These two BDT input variables are used as well in the photon-enriched sample, although their discrimination
power is limited given the $\onn>0.75$ requirement applied to both photons.}
\label{fig:onn}
\end{figure}
%%%%%%%%%%%%%%

Finally, the $\onn$ distributions for the leading photon ($O_{\rm NN}^{\gamma 1}$)
and the trailing photon ($O_{\rm NN}^{\gamma 2}$) show discrimination between
signal and the $\gamma j$ and $jj$ backgrounds, in particular in the jet-enriched
sample, as illustrated in Fig.~\ref{fig:onn}. The observed discrepancies between the
data and the total prediction in the shape of the distribution are partly covered by the 
combination of statistical uncertainties on the templates and the systematic uncertainties, 
and they have been checked to have a negligible impact on the final result.

\begin{figure}[t]
\centering
\includegraphics[width=0.45\textwidth]{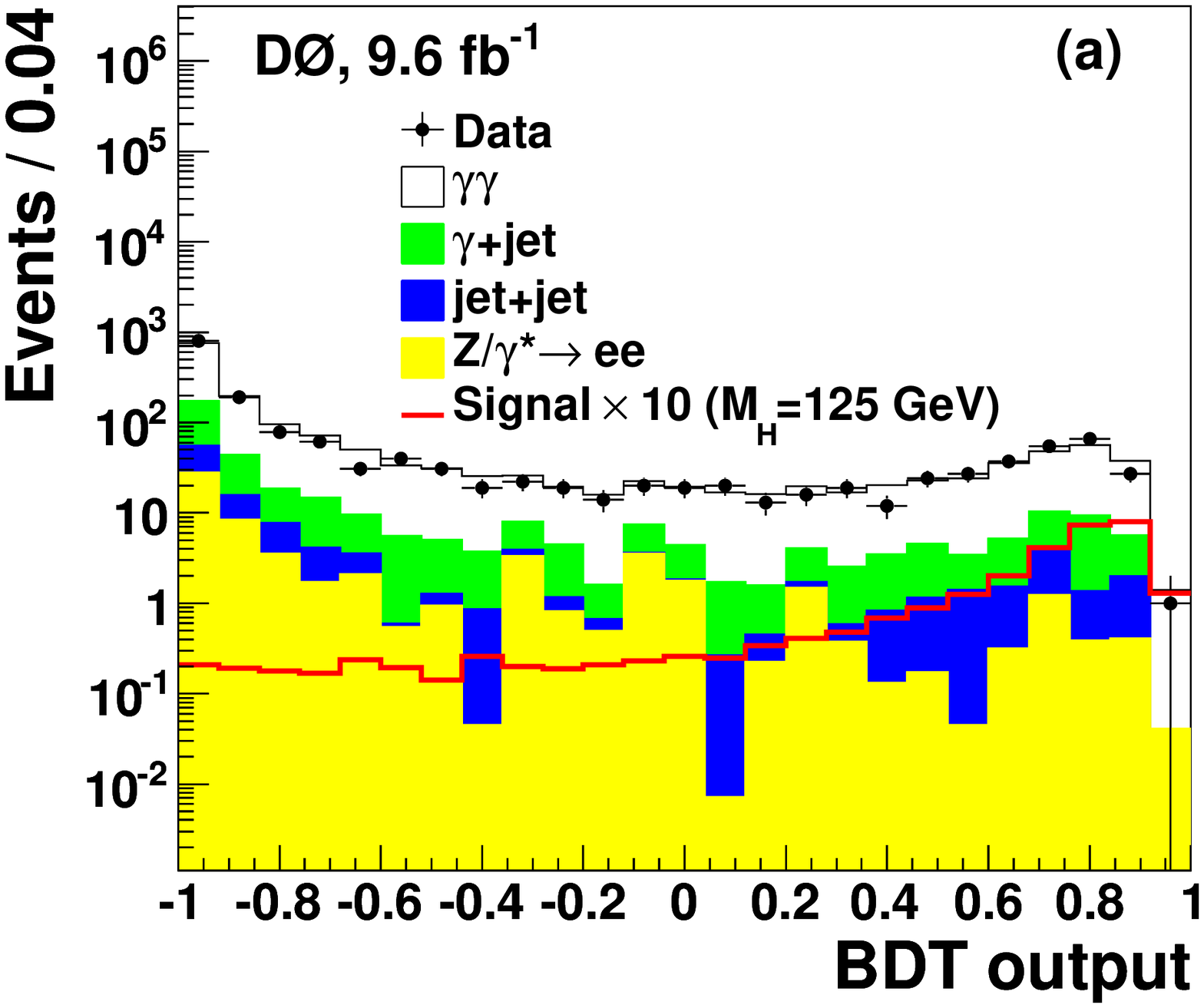}
\includegraphics[width=0.45\textwidth]{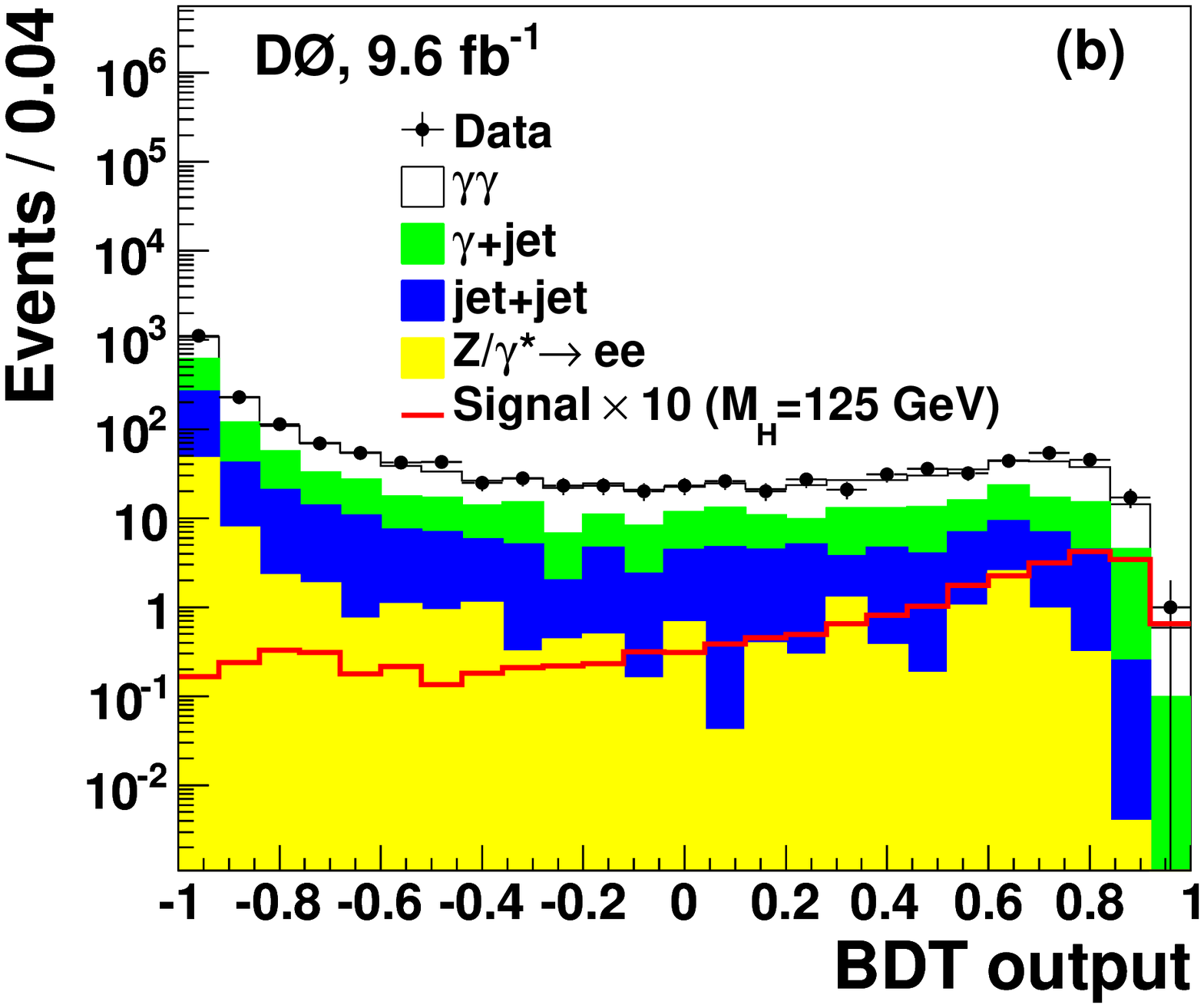}
\caption{\small (color online). Distribution of the BDT output  used in the SM Higgs boson search in (a) the photon-enriched sample and (b) the jet-enriched sample.
The data (points with statistical error bars) are compared to the background prediction, broken down into its individual components.
The expected distributions for a SM Higgs boson with $M_H=125\gev$ are also shown 
scaled by a factor of 10.}
\label{fig:bdt_sm}
\end{figure}

\begin{figure}[htbp]
\centering
\includegraphics[width=0.45\textwidth]{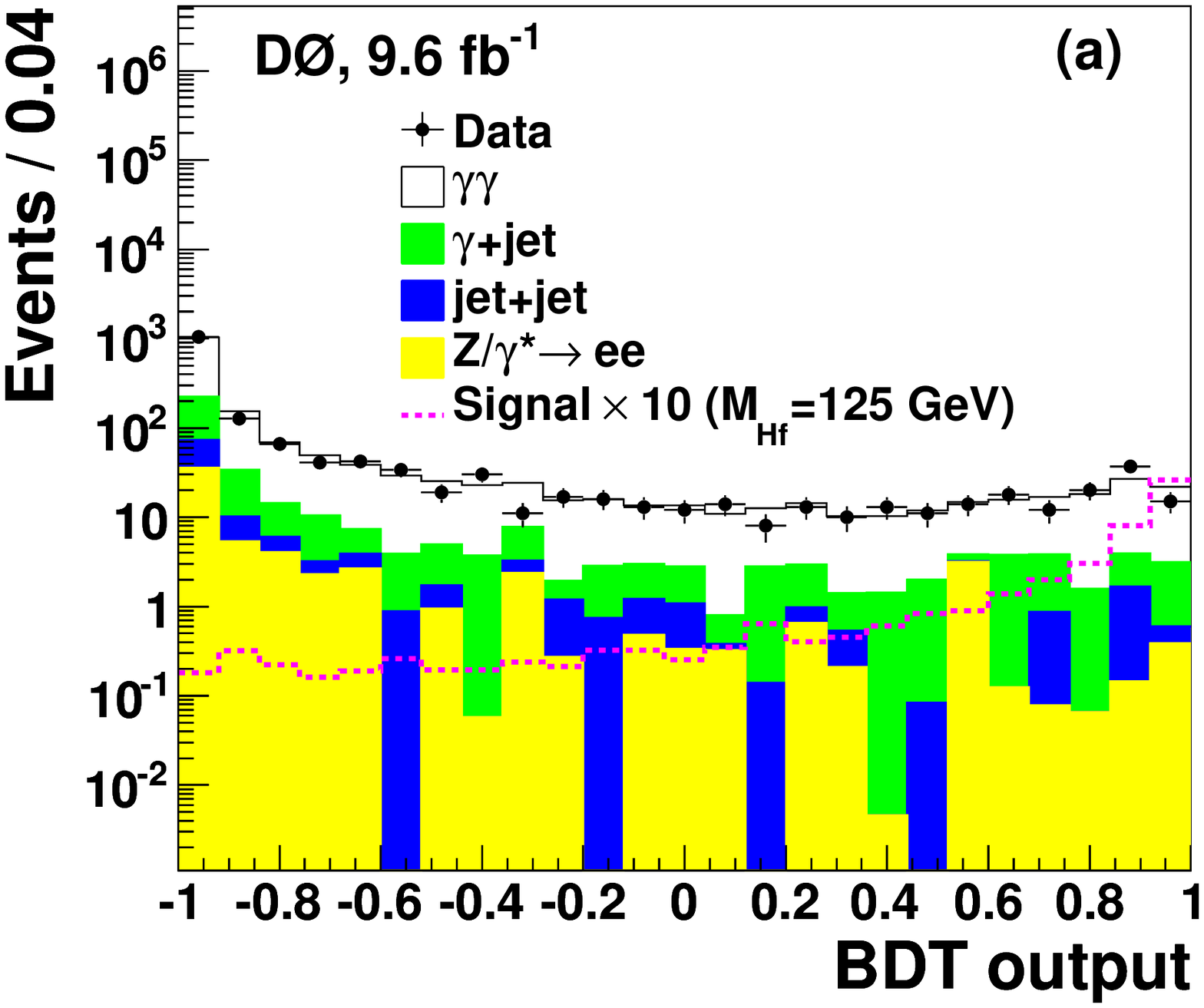}
\includegraphics[width=0.45\textwidth]{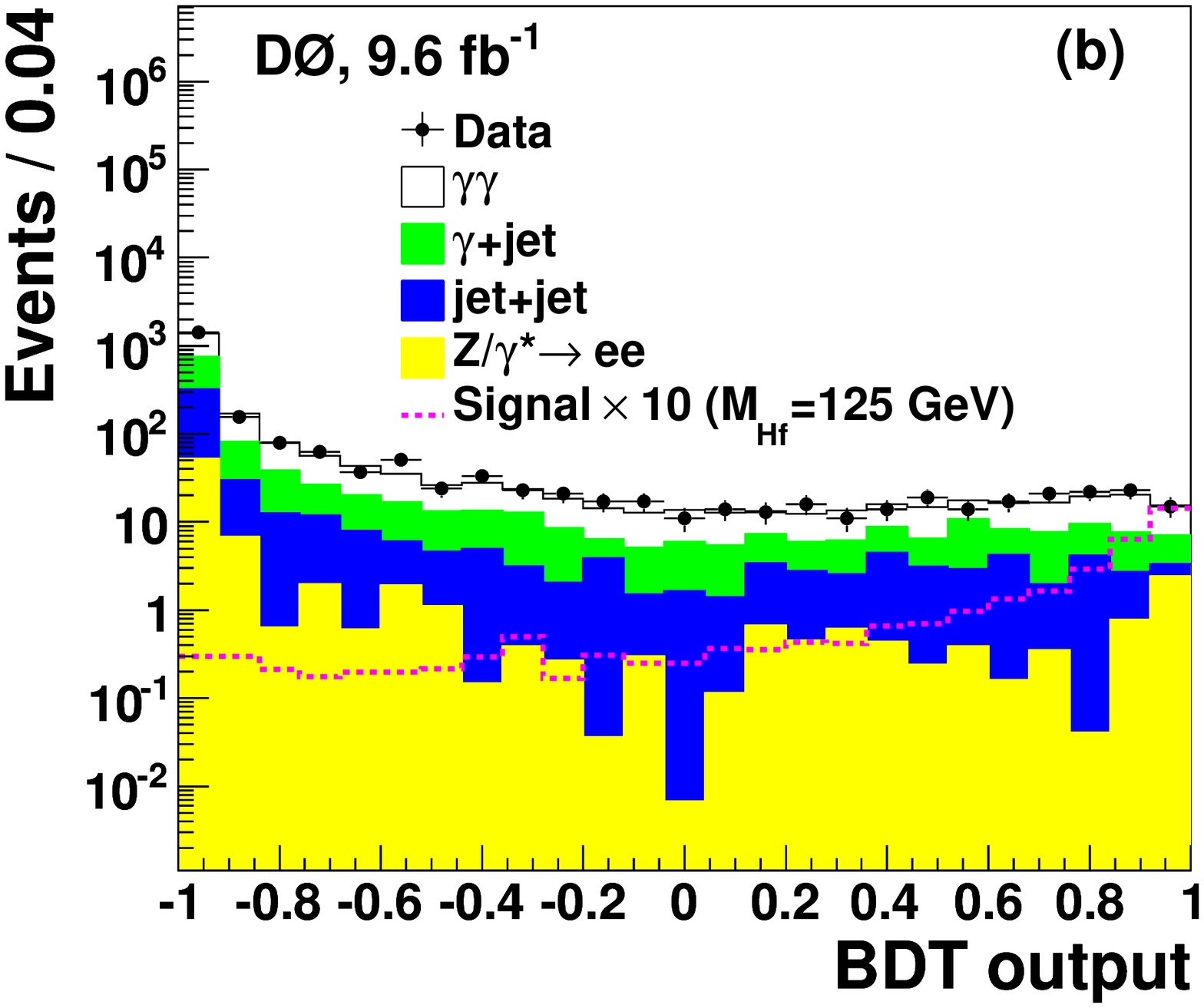}
\caption{\small (color online). Distribution of BDT output used in the fermiophobic  Higgs boson search in (a) the photon-enriched sample and (b) the jet-enriched sample.
The data (points with statistical error bars) are compared to the background prediction, broken down into its individual components.
The expected distributions for a fermiophobic Higgs boson with $M_{H_{\rm f}}=125\gev$ are also shown 
scaled by a factor of 10.}
\label{fig:bdt_fh}
\end{figure}

To improve the sensitivity of the search, a boosted-decision-tree (BDT)
technique~\cite{bdt} is used to build a single discriminating variable combining the
information from the ten variables.  A different BDT is trained, for each
$M_H$ hypothesis, for events selected in the search region, corresponding to
$\mgg$ falling in the interval of $M_H \pm 30\gev$.
The training is performed separately for the SM and the fermiophobic Higgs bosons
models, considering in each case the sum of all relevant signals against the
sum of all backgrounds. A separate BDT is trained in the photon-enriched and 
jet-enriched samples, respectively.  The resulting BDT output distributions assuming
a SM  and a fermiophobic Higgs boson with $M_H=125\gev$ are shown 
in Figs.~\ref{fig:bdt_sm} and~\ref{fig:bdt_fh}, respectively. Prior to fitting the background
yields to the data, these distributions are well modeled by the simulation and no significant excess above
the background prediction is observed at high values of the BDT output.

%%%%%%%%%%%%%%
%%%%%%%%%%%%%%

%%%%%%%%%%%%%%
%%%%%%%%%%%%%%

\section{Systematic uncertainties}

Systematic uncertainties affecting the normalization and shape of the BDT output distributions
are estimated for both signal and backgrounds, taking into account correlations.
Experimental uncertainties affecting the normalization of the signal and the $Z/\gamma^*\to e^+e^-$ background
include the integrated luminosity (6.1\%), tracking system live-time correction (2.0\%), trigger efficiency (0.1\%),  PV reconstruction efficiency (0.2\%),  
and photon identification efficiency for signal (3.9\%) or electron misidentification rate for $Z/\gamma^*\to e^+e^-$ (12.7\%). The impact from PDF uncertainties
on the signal acceptance is 1.7\%--2.2\% depending on $M_H$. Additional sources of uncertainty
affecting the normalization result from uncertainties on the theoretical cross section (including variations of the renormalization and factorization 
scales~\cite{signalscaleuncertainty} and the PDFs~\cite{pdfuncertainty}) for signal (GF (14.1\%), VH (6.2\%) and VBF (4.9\%)) and $Z/\gamma^*\to e^+e^-$ (3.9\%) production.

The normalization uncertainties affecting the $\gamma j$ and $jj$ background predictions result from
propagating the uncertainties on $\epsilon_{\gamma}$ (1.5\%) and $\epsilon_j$ (10\%) 
in the estimation of their yields via Eq.~\ref{matrix1}. The uncertainties on the $\gamma j$ and $jj$ yields
from varying $\epsilon_{\gamma}$ are 6.9\% and 5.3\%, respectively. The corresponding uncertainties from varying
$\epsilon_j$ are 0.6\% and 15.3\%, respectively.

The remaining systematic uncertainties affect the shape of the BDT output distributions.
Such uncertainties include the photon energy scale (1\%--5\% for signal, 1\%--4\% for DPP background), 
the modeling of DPP by {\sc sherpa} (1\%--10\%), and the modeling
of the Higgs boson $\pt$ spectrum in GF production (1\%--5\%). The last two uncertainties
are obtained by doubling and halving the factorization and renormalization scales with respect to the
nominal choice. Uncertainties on the shape of the $\gamma j+jj$ background are 5\%--7\% and are
estimated by comparing the BDT output distribution from the high-statistics samples obtained by inverting
the $\onn$ requirement to those predicted via the matrix method.

\section{Results}

For each hypothesized $M_H$ value, the BDT output distributions discussed in Sect.~\ref{sec:mva} for
the photon-enriched and jet-enriched samples are used to perform the statistical analysis to search 
for a significant signal above the background prediction. As mentioned before, such discriminants
are defined only for events with $\mgg$ falling in the $M_H \pm 30\gev$ interval.
The remainder of the $\mgg$ spectrum (see Fig.~\ref{fig:mgg}) for both the photon-enriched and jet-enriched
samples, corresponding to the sideband regions, is also included in the statistical analysis  as it 
provides a significant constraint on the DPP normalization. Therefore, for each $M_H$ a total of four distributions are analyzed.

\begin{figure*}[htbp]
\centering
\includegraphics[width=0.45\textwidth]{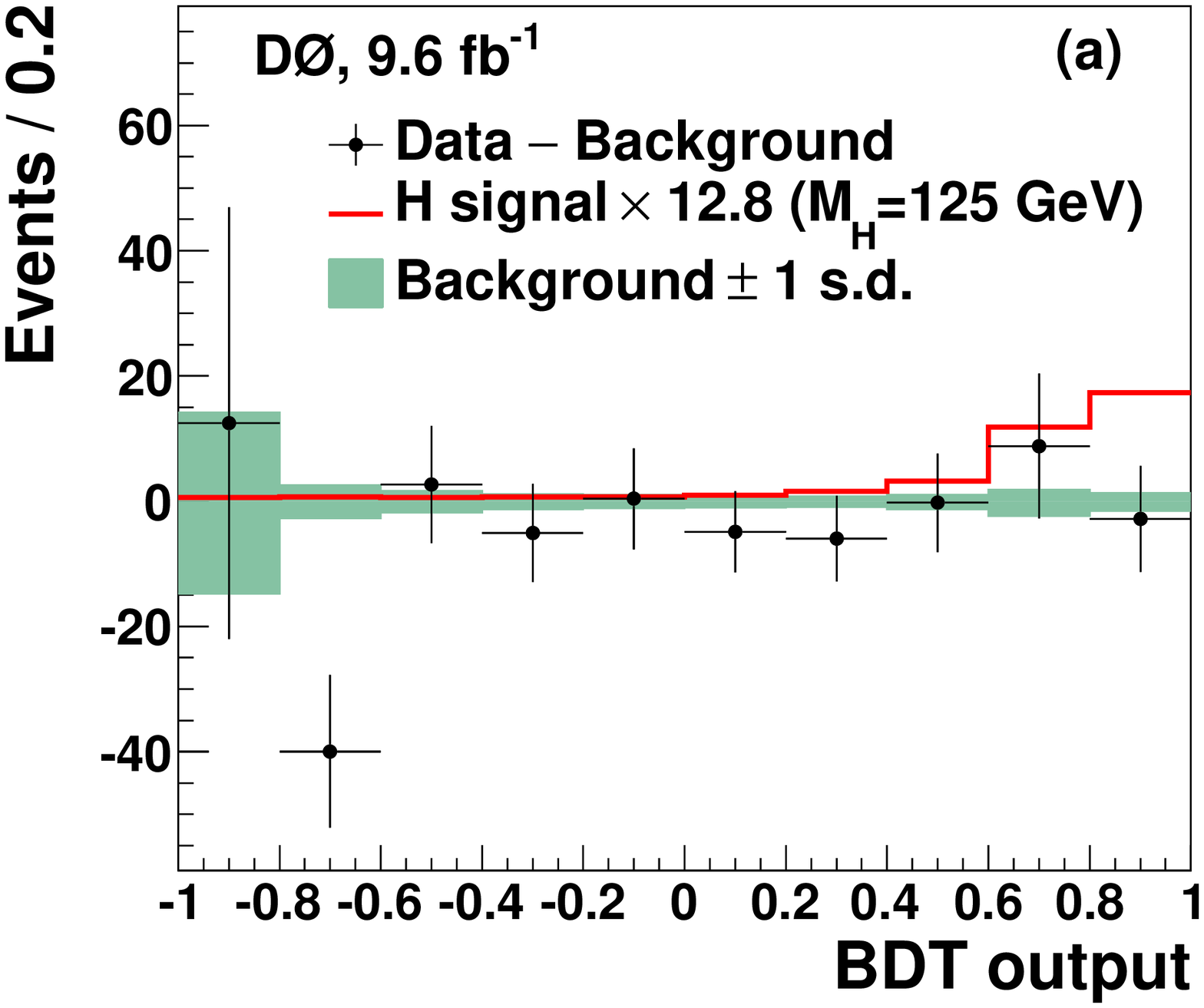}
\includegraphics[width=0.45\textwidth]{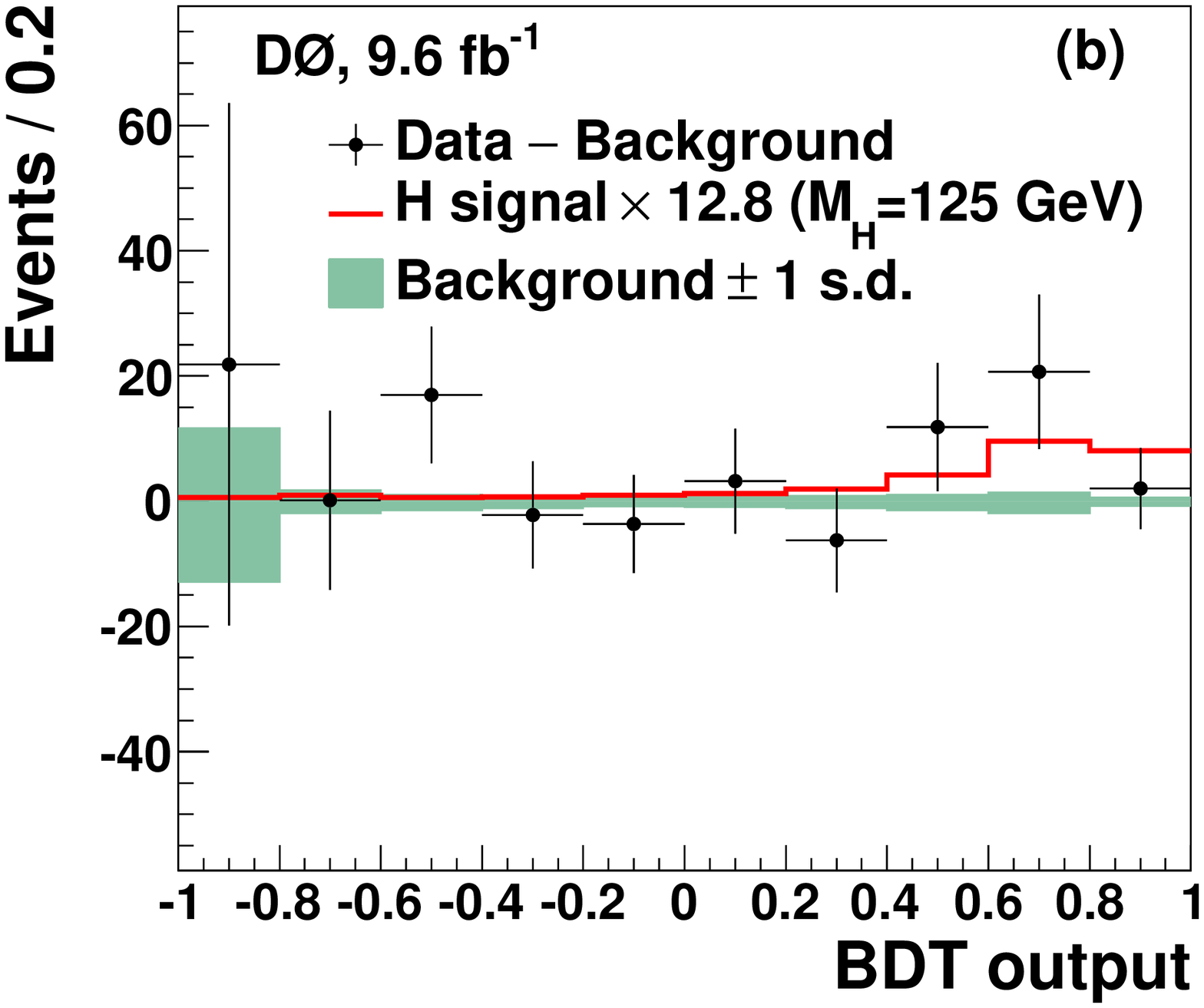}
\caption{\small (color online). 
Distribution of the BDT output for data (points with statistical error bars) after subtraction of the fitted background (under the background-only hypothesis) in (a) the photon-enriched sample and (b) the jet-enriched sample, for $M_H=125\gev$.
The expected SM Higgs signal is normalized to the observed limit on $\sigma \times {\cal B}(\Hgg)$. 
The bands represent the 1 s.d. uncertainties on the background prediction resulting from the fit.}
\label{fig:bdt_postfit}
\end{figure*}

In the absence of a significant data excess above the background prediction,  upper limits on the product of the 
production cross section and branching fraction ($\sigma \times {\cal B}(\Hgg)$) are derived
as a function of $M_H$, for both the SM and fermiophobic Higgs boson scenarios. 
Limits are calculated at the 95\% CL  with the modified frequentist approach~\cite{CLs-1}, which employs a log-likelihood ratio (LLR) as test-statistic,
${\rm LLR}=-2\ln (L_{\rm s+b}/L_{\rm b})$, where $L_{\rm s+b}$ ($L_{\rm b}$) is a binned likelihood function (product 
of Poisson probabilities) to observe the data under the signal-plus-background (background-only) hypothesis. 
Pseudo-experiments are generated for both hypotheses, taking into account per-bin statistical fluctuations
of the total predictions according to Poisson statistics, as well as Gaussian fluctuations describing the effect of systematic uncertainties.
The individual likelihoods are maximized with respect to the DPP background normalization 
as well as other nuisance parameters that parameterize the systematic uncertainties~\cite{CLs-2}.
This global fit determines the normalization of the DPP background directly from data 
and significantly reduces the impact of systematic uncertainties on the overall sensitivity.
Examples of the post-fit BDT output distribution, after background subtraction, are shown in Fig.~\ref{fig:bdt_postfit}.
The fraction of pseudo-experiments for the signal-plus-background (background-only) hypothesis with LLR larger than a given 
threshold defines ${\rm CL}_{\rm s+b}$ (${\rm CL}_{\rm b}$).  This threshold is set to the observed (median) LLR for the observed (expected) limit. 
Signal cross sections for which ${\rm CL}_{\rm s}={\rm CL}_{\rm s+b}/{\rm CL}_{\rm b}<0.05$ are deemed to be excluded at 95\% CL.

\begin{figure*}[htbp]
\centering
\includegraphics[width=0.45\textwidth]{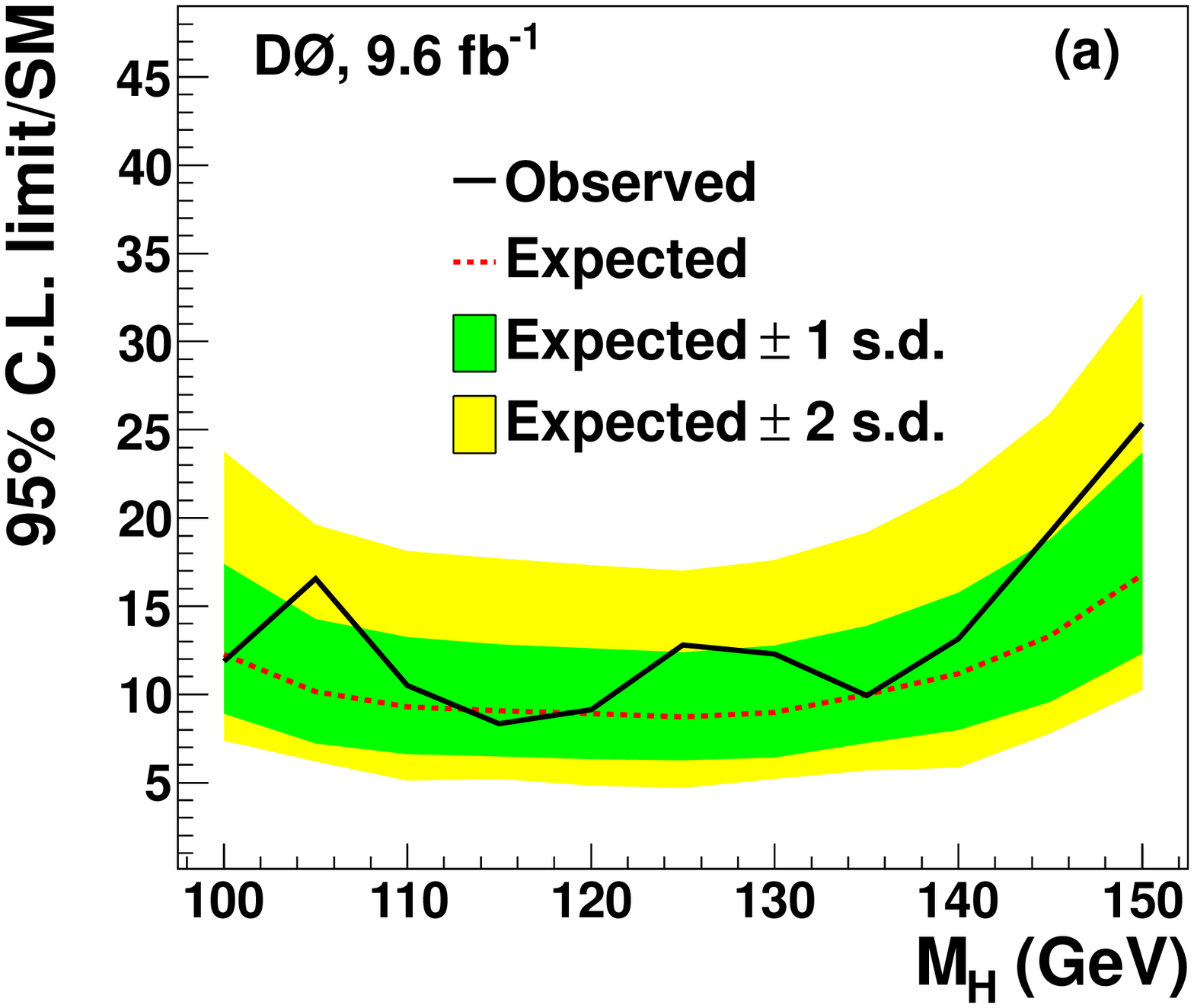}
\includegraphics[width=0.45\textwidth]{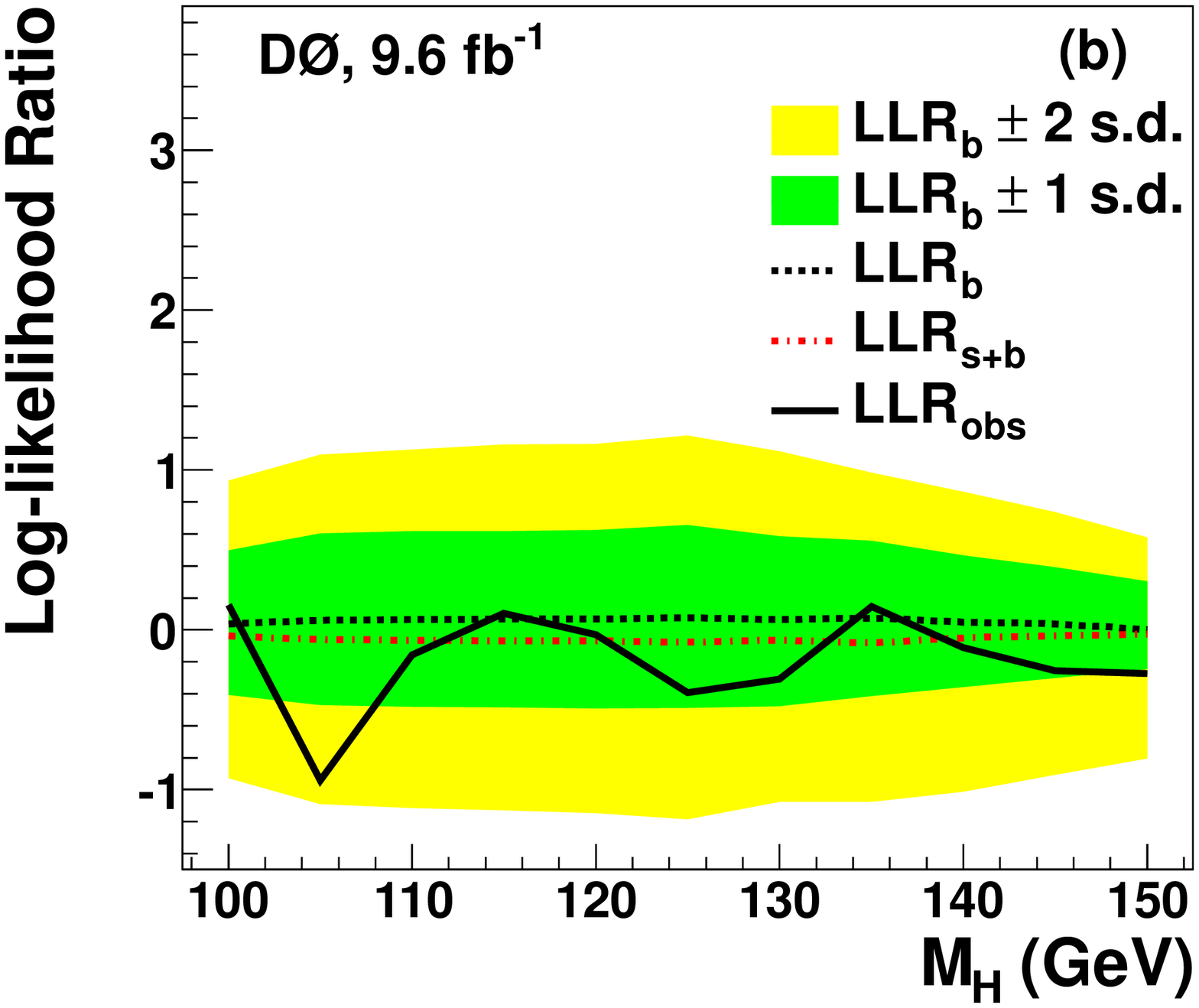}
\caption{\small (color online).
(a) Observed and expected 95\% CL limits on the ratio of $\sigma \times {\cal B} (H \to \gamma\gamma)$
to the SM prediction as a function of $M_H$. The bands correspond to 1 and 2 s.d. around the median expected limit under the background-only hypothesis.
(b) Observed log-likelihood ratio (LLR) as a function of $M_H$ compared to the expected LLR under the background-only hypothesis (LLR$_{\rm b}$) and signal+background hypothesis (LLR$_{\rm s+b}$). The bands correspond to the 1 s.d. and 2 s.d. around the expected median LLR$_{\rm b}$.}
\label{fig:limits_sm}
\end{figure*}

The resulting upper limits on $\sigma \times {\cal B}(\Hgg)$ relative to the SM prediction are shown 
as a function of $M_H$ in Fig.~\ref{fig:limits_sm}(a), and are summarized in Table~\ref{tab:limits_sm}, representing the most constraining results 
for a SM Higgs boson decaying into diphotons at the Tevatron. 
The corresponding LLR distribution is shown in Fig.~\ref{fig:limits_sm}(b).  The observed local excesses of data are
under 2 s.d. and therefore are consistent with background fluctuations. At $M_H=125\gev$ the best-fit
signal cross section is  a factor of $4.2\pm 4.6$ above the SM prediction. At the same mass, the value of ${\rm CL}_{\rm s+b}$ is 0.72 while
the p-value for the background-only hypothesis is $1-{\rm CL}_{\rm b}=0.20$.

\begin{figure*}[htbp]
\centering
\includegraphics[width=0.45\textwidth]{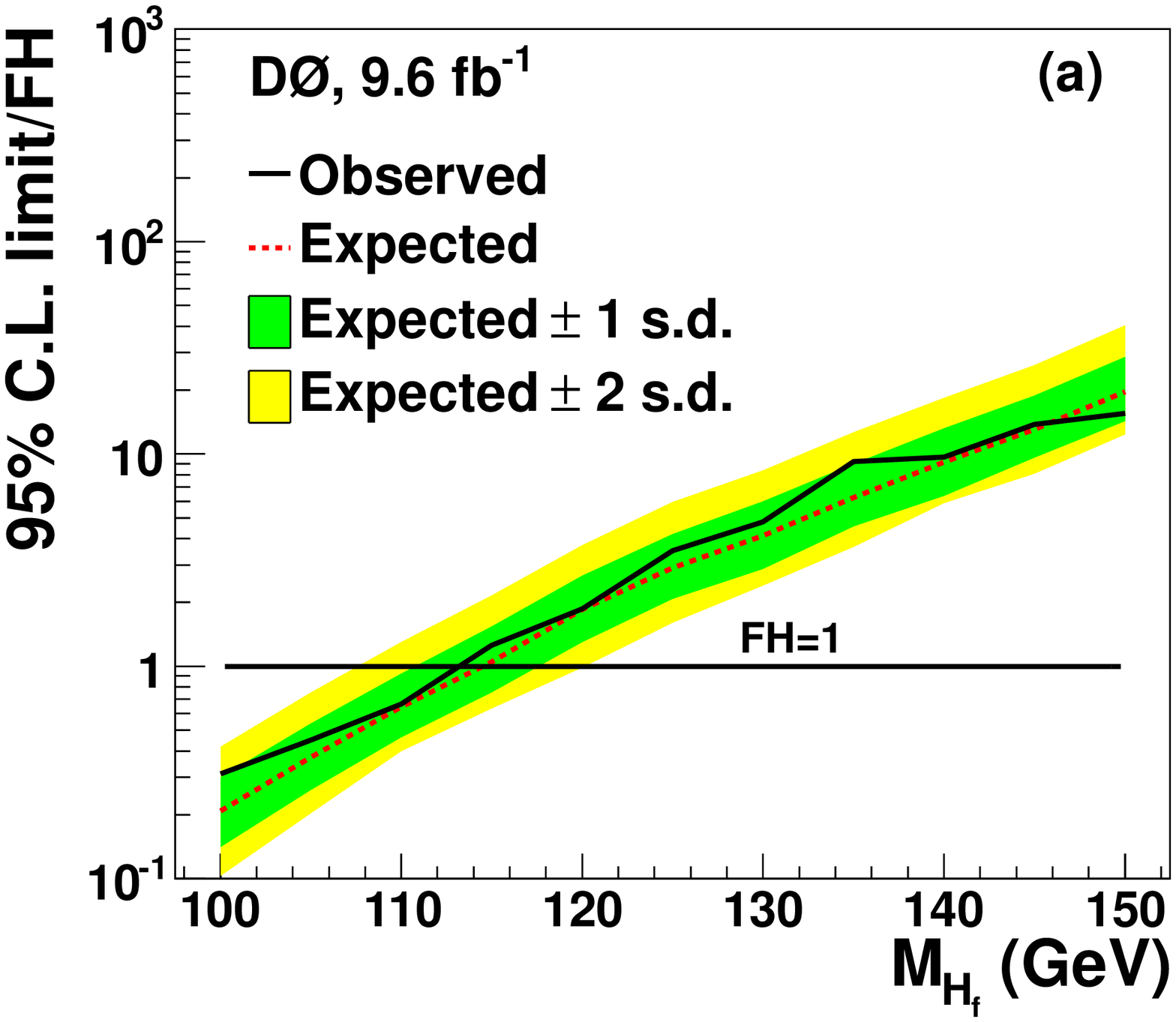}
\includegraphics[width=0.45\textwidth]{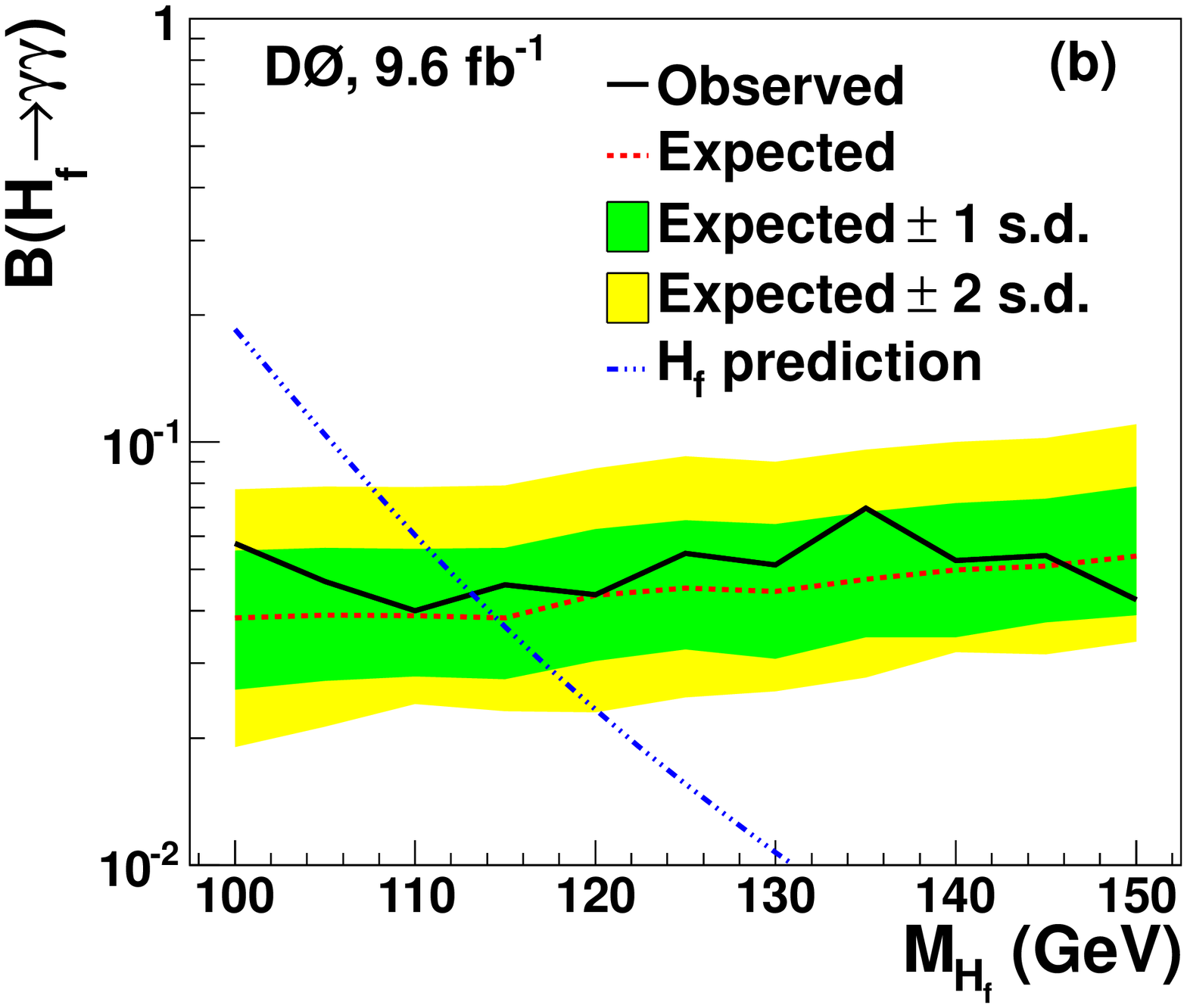}
\caption{\small (color online).
(a) Observed and expected 95\% CL limits on the ratio of $\sigma \times {\cal B} (H \to \gamma\gamma)$
to the fermiophobic Higgs model prediction as a function of $M_{H_{\rm f}}$. 
The bands correspond to 1 and 2 s.d. around the median expected limit under the background-only hypothesis.
(b) Observed and expected 95\% CL limits on ${\cal B} (H_{\rm f} \to \gamma\gamma)$ as a function of $M_{H_{\rm f}}$.
The bands correspond to the 1 and 2 s.d. around the median expected limit under the background-only hypothesis. 
Also shown is the prediction for a fermiophobic Higgs boson.}
\label{fig:limits_fh}
\end{figure*}

Upper limits on $\sigma \times {\cal B}(\Hgg)$ relative to the fermiophobic Higgs model prediction are shown 
as a function of $M_{H_{\rm f}}$ in Fig.~\ref{fig:limits_fh}(a), and are summarized in Table~\ref{tab:limits_fh}. This translates into the
observed (expected) lower 95\% CL of  $M_{H_{\rm f}} > 113$ (114)$\gev$.  After dividing by the theoretical cross section, upper limits on
${\cal B}(\Hfgg)$ are derived as a function of $M_{H_{\rm f}}$ and presented in Fig.~\ref{fig:limits_fh}(b). 

%%%%%%%%%%%%%%
%%%%%%%%%%%%%%

%%%%%%%%%%%%%%
%%%%%%%%%%%%%%

%%%%%%%%%%%%%%
%%%%%%%%%%%%%%

%%%%%%%%%%%%%%
\begin{table*}
\centering
\begin{tabular}{lcccccccccccc}
\hline \hline
 & $M_H$ (\gev) & 100 & 105 & 110 & 115 & 120 & 125 & 130 & 135 & 140 & 145 & 150 \\
\hline
$\sigma \times {\cal B}(\Hgg)$ (fb) & Expected  & 46.1& 37.2& 32.8& 30.3& 27.7& 24.6& 22.0& 20.7& 18.7& 17.2& 15.9 \\
		                                           & Observed & 44.7& 60.6& 37.1& 27.9& 28.4& 36.1& 30.1& 20.5& 22.0& 24.8& 24.0 \\ 
\hline   
$\sigma \times {\cal B}(\Hgg)$/SM  & Expected   & 12.2& 10.2& 9.3& 9.1& 8.9& 8.7& 9.0& 10.0& 11.2& 13.3& 16.8 \\
				                         & Observed   & 11.9& 16.6& 10.5& 8.3& 9.1& 12.8& 12.3& 9.9& 13.2& 19.2& 25.4 \\  				                         
\hline\hline 
\end{tabular}
\caption{\label{tab:limits_sm} Expected and observed upper limits at 95\% CL on the cross section times branching fraction for 
$\Hgg$ ($\sigma \times {\cal B}(\Hgg)$) and on $\sigma \times {\cal B}(\Hgg)$ relative to the SM prediction for a SM Higgs 
boson as a function of $M_H$.}
\end{table*}
%%%%%%%%%%%%%%

%%%%%%%%%%%%%%
\begin{table*}
\centering
\begin{tabular}{lcccccccccccc}
\hline \hline
 & $M_{H_{\rm f}}$ (\gev) & 100 & 105 & 110 & 115 & 120 & 125 & 130 & 135 & 140 & 145 & 150 \\
\hline
$\sigma \times {\cal B}(\Hfgg)$ (fb) & Expected  & 20.9 &18.3 &15.9 &13.7 &13.6 &12.4 &10.8 &10.2 &9.5 &8.6 &8.1 \\
                                                              & Observed  & 31.3 &22.0 &16.3 &16.4 &13.7 &15.0 &12.5 &15.0 &10.0 &9.1 &6.4 \\ 
		                                            & Theoretical prediction &100.4 &49.0 &24.7 &13.1 &7.3 &4.3 &2.6 &1.6 &1.0 &0.7 &0.4 \\
\hline
${\cal B}(\Hfgg)$ (\%) & Expected   & 3.8 &3.9 &3.9 &3.8 &4.3 &4.5 &4.4 &4.7 &5.0 &5.1 &5.4 \\
			           & Observed   & 5.8 &4.7 &4.0 &4.6 &4.4 &5.5 &5.1 &7.0 &5.3 &5.4 &4.2 \\  
	                              & Theoretical prediction  & 18.5 & 10.4 & 6.0 & 3.7 & 2.3 & 1.6 & 1.1 & 0.8 & 0.5 & 0.4 & 0.3 \\  
\hline\hline 
\end{tabular}
\caption{\label{tab:limits_fh}  Expected and observed upper limits at 95\% CL on the cross section times branching fraction for $H_{\rm f}\to\gamma\gamma$ ($\sigma \times {\cal B}(\Hfgg)$) and on ${\cal B}(\Hfgg)$ for a fermiophobic Higgs boson as a function of $M_{H_{\rm f}}$. Also given are the theoretical predictions for 
$\sigma \times {\cal B}(\Hfgg)$ and ${\cal B}(\Hfgg)$ as a function of $M_{H_{\rm f}}$. }
\end{table*}
%%%%%%%%%%%%%%

\section{Summary}
A search for a Higgs boson decaying into a pair of photons has been presented using 9.6 fb$^{-1}$ of $p\bar{p}$ collisions at $\sqrt{s} = 1.96\tev$ 
collected with the D0 detector at the Fermilab Tevatron Collider. The search employs multivariate
techniques to discriminate the signal from the non-resonant background, and is separately optimized
for a SM and a fermiophobic Higgs boson. No significant excess of data above the background prediction
is observed, and upper limits on the product of the cross section and branching fraction are derived at the 95\% CL 
as a function of $M_H$. For a SM Higgs boson with $M_H=125\gev$, the observed (expected) upper limits 
are a factor of 12.8 (8.7) above the SM prediction. The existence of a fermiophobic Higgs boson with mass in the 100--113$\gev$ 
range is excluded at the 95\% confidence level.

% acknowledgement.tex                            11 January 2013 
%
We thank the staffs at Fermilab and collaborating institutions,
and acknowledge support from the
DOE and NSF (USA);
CEA and CNRS/IN2P3 (France);
MON, NRC KI and RFBR (Russia);
CNPq, FAPERJ, FAPESP and FUNDUNESP (Brazil);
DAE and DST (India);
Colciencias (Colombia);
CONACyT (Mexico);
NRF (Korea);
FOM (The Netherlands);
STFC and the Royal Society (United Kingdom);
MSMT and GACR (Czech Republic);
BMBF and DFG (Germany);
SFI (Ireland);
The Swedish Research Council (Sweden);
and
CAS and CNSF (China).


\begin{thebibliography}{50}
\bibitem{sm}
S.L.~Glashow, Nucl. Phys. {\bf 22}, 579 (1961);
S.~Weinberg, Phys. Rev. Lett. {\bf 19}, 1264 (1967);
A.~Salam, Proceedings of the 8th Nobel Symposium, ed. N. Svartholm, Almquist and Wiksel, Stockholm, 367 (1968).

\bibitem{higgs}
F.~Englert and R.~Brout, Phys. Rev. Lett. {\bf 13}, 321 (1964);
P.W.~Higgs, Phys. Rev. Lett. {\bf 13}, 508 (1964);
P.W.~Higgs, Phys. Lett. {\bf 12}, 132 (1964);
G.~Guralnik, C.~Hagen, and T.~Kibble, Phys. Rev. Lett. {\bf 13}, 585 (1964).

\bibitem{lepewwg}
The LEP Electroweak Working Group, ``Status of March 2012",
\url{http://lepewwg.web.cern.ch/LEPEWWG/}.

\bibitem{lepcombo}
ALEPH, DELPHI, L3, and OPAL Collaborations, LEP Working Group for Higgs Boson Searches, 
Phys. Lett. B {\bf 565}, 61 (2003).

\bibitem{tevcombo}
Tevatron New Physics and Higgs Working Group and CDF and D0 Collaborations, arXiv:1207.0449 [hep-ex].

\bibitem{tevcombobb}
T.~Aaltonen {\sl et al.} (CDF and D0 Collaborations), Phys. Rev. Lett. {\bf 109}, 071804 (2012);
T.~Aaltonen {\sl et al.} (CDF Collaboration), Phys. Rev. Lett. {\bf 108}, 151803 (2012);
V.M.~Abazov {\sl et al.} (D0 Collaboration), Phys. Rev. Lett. {\bf 108}, 151804 (2012).

\bibitem{atlascombo}
ATLAS Collaboration, Phys. Lett. B {\bf 716}, 1 (2012).

\bibitem{cmscombo}
CMS Collaboration, Phys. Lett. B {\bf 710}, 26 (2012); Phys. Lett. B {\bf 716}, 30 (2012).

\bibitem{beyond-SM}
S. Mrenna and J. Wells, Phys. Rev. D {\bf 63}, 015006 (2000).

\bibitem{Plehn}
T.~Plehn and M.~Rauch, Europhys. Lett. {\bf 100}, 11002 (2012).

\bibitem{LEP-FH}
A.~Heister {\sl et al.} (ALEPH Collaboration), Phys. Lett. B {\bf 544}, 16 (2002);
P.~Abreu {\sl et al.} (DELPHI Collaboration),  Eur. Phys. J. C {\bf 35}, 313 (2004);
P.~Achard {\sl et al.} (L3 Collaboration), Phys. Lett. B {\bf 568}, 191 (2003);
G.~Abbiendi {\sl et al.} (OPAL Collaboration), Phys. Lett. B {\bf 544}, 44 (2002);
A.~Rosca (LEP Collaborations), arXiv:hep-ex/0212038.

\bibitem{CDF-Hgg}
T.~Aaltonen {\sl et al.} (CDF Collaboration), Phys. Lett. B {\bf 717}, 173 (2012).

\bibitem{D0-Hgg}
V.M.~Abazov {\sl et al.} (D0 Collaboration), Phys. Rev. Lett. {\bf 107}, 151801 (2011).

\bibitem{ATLAS-Hgg-FH}
ATLAS Collaboration, Eur. Phys. J. C {\bf 72}, 2157 (2012).

\bibitem{CMS-Hgg-FH}
CMS Collaboration, J. High Energy Phys.  {\bf 1209}, 111 (2012).

\bibitem{d0det}
V.~M.~Abazov {\sl et al.}, Nucl. Instrum. Methods Phys. Res. A {\bf 565}, 463 (2006);
S.~N.~Ahmed {\sl et al.}, Nucl. Instrum. Methods Phys. Res. A {\bf 634}, 8 (2011);
M.~Abolins {\sl et al.}, Nucl. Instrum. Methods Phys. Res. A {\bf 584}, 75  (2008);
R.~Angstadt {\sl et al.}, Nucl. Instrum. Methods Phys. Res. A {\bf 622}, 298  (2010).

\bibitem{d0_coordinate}
Pseudorapidity is defined as $\eta=-\ln[\tan(\theta/2)]$, where $\theta$ is the polar angle
relative to the proton beam direction, and $\phi$ is the azimuthal angle in the plane transverse 
to the proton beam direction.

\bibitem{d0lumi}
T.~Andeen {\sl et al.}, FERMILAB-TM-2365 (2007).

\bibitem{pythia}
T.~Sj\"ostrand {\sl et al.}, J. High Energy Phys. {\bf 05}, 026 (2006).
Version 6.409 is used.

\bibitem{cteq}
J.~Pumplin {\sl et al.}, J. High Energy Phys. {\bf 07}, 012 (2002);
D.~Stump {\sl et al.}, J. High Energy Phys. {\bf 10}, 046 (2003).

\bibitem{ggHxsect}
C.~Anastasiou, R.~Boughezal, and F.~Petriello, J. High Energy Phys. {\bf 04}, 003 (2009);
D.~Florian and M.~Grazzini, Phys. Lett. B {\bf 674}, 291 (2009).

\bibitem{VHxsect}
J.~Baglio and A.~Djouadi, J. High Energy Phys. {\bf 10}, 064 (2010).

\bibitem{VBFxsect}
P.~Bolzoni, F.~Maltoni, S.O.~Moch, and M.~Zaro, Phys. Rev. Lett. {\bf 105}, 011801 (2010).

\bibitem{signalPDF}
A.D.~Martin, W.J.~Stirling, R.S.~Thorne and G.~Watt, Eur. Phys. J. C {\bf 63}, 189 (2009).

\bibitem{hdecay}
A.~Djouadi, J.~Kalinowski, and M.~Spira, Comput. Phys. Commun. {\bf 108}, 56 (1998).
Version 3.70 is used.

\bibitem{hqt}
G.~Bozzi, S.~Catani, D.~de Florian, and M.~Grazzini, Phys. Lett. B {\bf 564}, 65 (2003);
Nucl. Phys. {\bf B737}, 73 (2006).

\bibitem{sherpa}
T.~Gleisberg {\sl et al.}, J. High Energy Phys. {\bf 02}, 007 (2009).
Version 1.2.2 is used.

\bibitem{diphotonPLB}
V.M.~Abazov {\sl et al.}, (D0 Collaboration), arXiv:1301.4536 [hep-ex], submitted to Phys. Lett. B.

\bibitem{alpgen}
M.L.~Mangano {\sl et al.}, J. High Energy Phys. {\bf 07}, 001 (2003).
Version 2.11 is used.

\bibitem{Zqt}
V.M.~Abazov {\sl et al.}, (D0 Collaboration), Phys. Rev. Lett. {\bf 100}, 102002 (2008).

\bibitem{Zxsec}
R.~Hamberg, W.L.~van Neerven and T. Matsuura, Nucl. Phys. B {\bf 359}, 343 (1991) [Erratum-ibid. B {\bf 644}, 403 (2002)].

\bibitem{geant}
R.~Brun and F.~Carminati, CERN Program Library Long Writeup W5013, 1993.

\bibitem{HOR}
V.M.~Abazov {\sl et al.} (D0 Collaboration), Phys. Lett. B {\bf 659}, 856 (2008).

\bibitem{ONN}
V.M.~Abazov {\sl et al.} (D0 Collaboration), Phys. Rev. Lett. {\bf 102}, 231801 (2009).

\bibitem{Zg}
V.M.~Abazov {\sl et al.} (D0 Collaboration), Phys. Lett. B {\bf 653}, 378 (2007).

\bibitem{bidnim}
V.M.~Abazov {\sl et al.} (D0 Collaboration), Nucl. Instrum. Methods Phys. Res. A {\bf 620}, 490 (2010).

\bibitem{kalman}
R.E.~Kalman, Trans. ASME (J. Basic Engineering), {\bf 82} D, 35 (1960).

\bibitem{pointing}
S.~Kesisoglou, Brown University, Ph.D. Thesis,
FERMILAB-THESIS-2004-44, UMI-31-74625, 2004.

\bibitem{CrystalBall}
J.E.~Gaiser, Ph.D. Thesis, SLAC-R-255 (1982).

\bibitem{bkg-subtract}
D.~Acosta {\sl et al.} (CDF Collaboration), Phys. Rev. Lett. {\bf 95}, 022003 (2005).

\bibitem{diphoton-Xsection}
V.M.~Abazov {\sl et al.}, (D0 Collaboration), Phys. Lett. B {\bf 690}, 108 (2010).

\bibitem{collins-soper-frame}
J.C. Collins and D.E. Soper, Phys. Rev. D {\bf 16}, 2219 (1977).

\bibitem{jets}
G.C.~Blazey {\sl et al.}, arXiv:hep-ex/0005012 (2000);
V.M.~Abazov {\sl et al.} (D0 Collaboration), Phys. Rev. D {\bf 85}, 052006 (2012).
Calorimeter jets are reconstructed using the iterative midpoint cone
algorithm with radius ${\cal R}=0.5$ in $y$--$\phi$ space,
where $y$ is the rapidity.  

\bibitem{bdt}
A.~Hoecker {\sl et al.}, arXiv:physics/0703039 [physics.data-an];
A.~Hoecker {\sl et al.}, PoS (ACAT) {\bf 040} (2007).

\bibitem{signalscaleuncertainty}
I.W.~Stewart and F.J.~Tackmann, Phys. Rev. D {\bf 85}, 034011 (2012).

\bibitem{pdfuncertainty}
S.~Alekhin, S.~Alioli, R.D.~Ball {\sl et al.}, arXiv:1101.0536 [hep-ph];
M.~Botje, J.~Butterworth, A.~Cooper-Sarkar {\sl et al.}, arXiv:1101.0538 [hep-ph].

\bibitem{CLs-1}
T.~Junk, Nucl. Instrum. Methods Phys. Res. A {\bf 434}, 435 (1999); A.~Read, J. Phys. G {\bf 28}, 2693 (2002).

\bibitem{CLs-2}
W.~Fisher, FERMILAB-TM-2386-E (2006).


\end{thebibliography}
\end{document}